\documentclass[twocolumn]{aastex62} 

\newcommand{\alw}{a$_{<LW>}$}
\newcommand{\amw}{a$_{<MW>}$}
\newcommand{\amed}{a$_{50}$}
\newcommand{\aft}{a$_{10}$}
\newcommand{\mfit}{M$_{*,FIT}$}
\newcommand{\lfit}{L$_{FIT}$}
\newcommand{\ebv}{$E(B-V)_i$}
\newcommand{\sigs}{$\sigma_*$}

\accepted{May 6, 2018}

\submitjournal{ApJ} 

\shorttitle{SFHs OF $Z\sim1$ GALAXIES IN LEGA-C}
\shortauthors{Chauke et al.}

\begin{document}

\title{STAR FORMATION HISTORIES OF $Z\sim1$ GALAXIES IN LEGA-C}

\email{chauke@mpia-hd.mpg.de}

\author[0000-0002-0786-7307]{Priscilla Chauke}
\affil{Max-Planck-Institut f\"ur Astronomie, K\"onigstuhl 17, D-69117 Heidelberg, Germany}

\author{Arjen van der Wel}
\affiliation{Sterrenkundig Observatorium, Universiteit Gent, Krijgslaan 281 S9, B-9000 Gent, Belgium}
\affiliation{Max-Planck-Institut f\"ur Astronomie, K\"onigstuhl 17, D-69117 Heidelberg, Germany}

\author{Camilla Pacifici}
\affiliation{Space Telescope Science Institute, 3700 San Martin Drive, Baltimore, MD 21218, USA}

\author{Rachel Bezanson}
\affiliation{University of Pittsburgh, Department of Physics and Astronomy, 100 Allen Hall, 3941 O'Hara St, Pittsburgh PA 15260, USA}

\author{Po-Feng Wu}
\affiliation{Max-Planck-Institut f\"ur Astronomie, K\"onigstuhl 17, D-69117 Heidelberg, Germany}

\author{Anna Gallazzi}
\affiliation{INAF-Osservatorio Astrofisico di Arcetri, Largo Enrico, Fermi 5, I-50125 Firenze, Italy}

\author{Kai Noeske}
\affiliation{Experimenta Heilbronn, Kranenstra{\ss}e 14, 74072, Heilbronn, Germany}

\author{Caroline Straatman}
\affiliation{Sterrenkundig Observatorium, Universiteit Gent, Krijgslaan 281 S9, B-9000 Gent, Belgium}

\author{Juan-Carlos Mu\~{n}os-Mateos}
\affiliation{European Southern Observatory, Alonso de Cordova 3107, Casilla 19001, Vitacura, Santiago, Chile}

\author{Marijn Franx}
\affiliation{Leiden Observatory, Leiden University, PO Box 9513, 2300 RA Leiden, The Netherlands}

\author{Ivana Bari\v{s}i\'{c}}
\affiliation{Max-Planck Institut f\"{u}r Astronomie, K\"{o}nigstuhl 17, D-69117, Heidelberg, Germany}

\author{Eric F. Bell}
\affiliation{Department of Astronomy, University of Michigan, 1085 S. University Ave., Ann Arbor, MI 48109, USA}

\author{Gabriel B. Brammer}
\affiliation{Space Telescope Science Institute, 3700 San Martin Drive, Baltimore, MD 21218, USA}

\author{Joao Calhau}
\affiliation{Physics Department, Lancaster University, Lancaster LA1 4YB, UK}

\author{Josha van Houdt}
\affiliation{Max-Planck Institut f\"{u}r Astronomie, K\"{o}nigstuhl 17, D-69117, Heidelberg, Germany}

\author{Ivo Labb\'{e}}
\affiliation{Leiden Observatory, Leiden University, PO Box 9513, 2300 RA Leiden, The Netherlands}

\author{Michael V. Maseda}
\affiliation{Leiden Observatory, Leiden University, PO Box 9513, 2300 RA Leiden, The Netherlands}

\author{Adam Muzzin}
\affiliation{Department of Physics and Astronomy, York University, 4700 Keele St., Toronto, Ontario, MJ3 1P3, Canada}

\author{Hans-Walter Rix}
\affiliation{Max-Planck Institut f\"{u}r Astronomie, K\"{o}nigstuhl 17, D-69117, Heidelberg, Germany}

\author{David Sobral}
\affiliation{Leiden Observatory, Leiden University, PO Box 9513, 2300 RA Leiden, The Netherlands}
\affiliation{Physics Department, Lancaster University, Lancaster LA1 4YB, UK}




\begin{abstract}
Using high resolution spectra from the VLT LEGA-C program, we reconstruct the star formation histories (SFHs) of 607 galaxies at redshifts $z = 0.6-1.0$ and stellar masses $\gtrsim10^{10}$\,M$_{\odot}$ using a custom full spectrum fitting algorithm that incorporates the \textit{emcee} and \textit{FSPS} packages. We show that the \added{mass-weighted} age of a galaxy correlates strongly with stellar velocity dispersion ($\sigma_*$) and ongoing star-formation (SF) activity, with the stellar content in higher-$\sigma_*$ galaxies having formed earlier and faster.  The SFHs of quiescent galaxies are generally consistent with passive evolution since their main SF epoch, but a minority show clear evidence of a rejuvenation event in their recent past. The mean age of stars in galaxies that are star-forming is generally significantly younger, with SF peaking after $z<1.5$ for almost all star-forming galaxies in the sample: many of these still have either constant or rising SFRs on timescales $>100$\,Myrs. This indicates that $z>2$ progenitors of $z\sim1$ star-forming galaxies are generally far less massive. Finally, despite considerable variance in the individual SFHs, we show that the current SF activity of massive galaxies ($>$\,L$_*$) at  $z\sim1$ correlates with SF levels at least $3$\,Gyrs prior: SFHs retain `memory' on a large fraction of the Hubble time. Our results illustrate a novel approach to resolve the formation phase of galaxies, and in identifying their individual evolutionary paths, connects progenitors and descendants across cosmic time. This is uniquely enabled by the high-quality continuum spectroscopy provided by the LEGA-C survey.
\end{abstract}




\section{Introduction}
\label{sec:intro}

The ability to reconstruct the star-formation histories of galaxies, by characterising their stellar populations, allows one to trace their individual evolution through time, and thereby directly connect their descendants to their progenitors at higher redshifts. Thus far, high-redshift galaxy surveys have produced snapshots of the galaxy population at different points in cosmic time, which produces tight boundary conditions for galaxy formation models. However, the importance of the many physical processes included in these models are not directly constrained. We still do not know individual star-formation histories (SFHs) and how these are related to global galaxy properties. To constrain galaxy formation theories more directly, `archaeological' reconstruction can be used to trace the evolution of individual galaxies over time, and then the dependance of individual SFHs on stellar mass, stellar velocity dispersion and star-formation (SF) activity can be explored.

Reconstructing SFHs requires high resolution spectra of galaxies. Ideally, individual stars would be resolved, as they are for local dwarf galaxies \citep[e.g.,][]{weisz2011}. However, in most cases we have to rely on integrated stellar light, though if a galaxy's main star formation (SF) epoch lies at $z>1$, we cannot temporally resolve its stellar age distribution, even with the highest-quality spectra. While there is a plethora of high resolution spectra of galaxies in the local universe, most of these galaxies are too old \citep[$>5$\,Gyrs,][]{gallazzi2005} to resolve their star-formation histories (SFHs) due to the similarity of stellar spectra in the age range $>5$\,Gyrs. The general insight gained from the `archaeological' studies of these galaxies is that low-mass galaxies have more extended SFHs that peak at later cosmic times compared to high-mass galaxies \citep[`downsizing', e.g.,][]{gallazzi2005,thomas2005,thomas2010}. \added{Many of these studies involved the use of fossil record methods on SDSS \citep[Sloan Digital Sky Survey,][]{york2000} spectra of local galaxies \citep[e.g. ][]{juneau2005, thomas2005, fernandes2007, tojeiro2009, mcdermid2015, ibarra2016}. However, downsizing has also been seen in other studies, such as studies by \cite{cimatti2006}, who corrected luminosity function data of early-type galaxies by adopting the empirical luminosity dimming rate derived from the evolution of the Fundamental Plane of field and cluster massive early-type galaxies, as well as \cite{leitner2012}, who derived the average growth of stellar mass in local star-forming galaxies using a Main Sequence Integration approach.}

One approach to probe the high-redshift regime, is to obtain an integrated view of galaxy evolution. Thus far, this has been the focus of spectroscopic observations of distant galaxies: the evolution of the star-formation rate density (SFRD) in the universe has been extensively studied \citep[e.g.,][]{karim2011, madau2014, khostovan2015, abramson2016}. The majority of these studies indicate that the SFRD increased from high redshift to $z\sim2$, and has since been decreasing steeply. Coupled with this, are number density evolution studies which show an increasingly dominant population of quiescent galaxies \citep[e.g.,][]{pozzetti2010, bram2011, moustakas2013, muzzin2013g}.

Another approach is to use photometric measurements to trace SFHs, however, individual galaxy evolution is not easily traced with this method due to high uncertainties. In this case, one can investigate average SFHs of galaxies as \citet{pacifici2016} have done by applying spectral energy distribution models to compute the median SFHs of 845 quiescent galaxies at $0.2< z< 2.1$. They found that galaxy stellar mass is a driving factor in determining how evolved galaxies are, with high mass galaxies being the most evolved at any time. The limitation with these approaches is that we cannot connect progenitors to descendants: studies from mass-matched samples have resulted in multiple solutions \cite[e.g.,][]{torrey2017}. To understand the mechanics of how galaxies evolve, it is crucial to expand our view from focusing on the population of galaxies as a whole, to investigating how the star-formation rate (SFR) of individual galaxies varies with time.

Probing the SFHs of individual galaxies, however, still requires high-resolution, high-quality stellar continuum spectra, which are expensive to obtain. Consequently, high-redshift samples are small and often selected with criteria to optimize data quality and sample size rather than represent the full galaxy population. \citet{jorgensen2013} obtained spectra for $\sim80$ cluster galaxies at $z=0.5-0.9$ and found ages of $3-6$~Gyrs, consistent with passive evolution between $z\sim2$ and the present. Stellar population measurements of $\sim70$ $z\sim0.7$ galaxies with stellar masses $>10^{10}$\,M$_{\odot}$ were performed by \cite{gallazzi2014}; they found that passive galaxies have ages and metallicities consistent with those of present-day galaxies, and that star-forming galaxies require further star-formation and metal enrichment to evolve into present-day descendants. \citet{choi2014} analysed stacked spectra of thousands of passive galaxies in the redshift range $0.1<z<0.7$ and also found age evolution consistent with mostly passive evolution, with little dependence on mass at $z>0.5$. \citet{belli2015} measured ages of $1-4$~Gyrs for several dozen passive galaxies at redshifts $1<z<1.6$, indicating that we are approaching the cosmic epoch at which massive, passive galaxies stopped forming stars. Finally, at $z>1.5$, measurements are limited to stacked spectra \citep[e.g.,][]{whitaker2013,onodera2015} or sample sizes ranging from single objects to a handful \citep[e.g.,][]{vandokkum2010,toft2012,vandesande2013,kriek2016}. The typical age of massive, passive galaxies at those redshifts is found to be 1 Gyr or less, with short formation time scales. From this brief review, it is evident that samples at large lookback time are generally small and/or stacked. Furthermore, ages are usually estimated by assuming a single stellar population, which is arguably justified for very massive galaxies at late cosmic epochs, but not in general.

The LEGA-C  \citep[Large Early Galaxy Astrophysics Census,][]{vdw2016} survey is collecting high $S/N$ spectra of $\sim3000$ galaxies in the redshift range $0.6<$ z $<1$, selected only by their $K$-band magnitude (a proxy for stellar mass). The data, which are comparable in quality to those obtained in the nearby universe, probe the internal kinematics of stars and gas, and the ages and metallicities of stellar populations. This enables us, for the first time, to reconstruct the SFHs of individual galaxies at large look-back time that are representative of the population. \deleted{The goal of this paper is to resolve the main formation phase of massive galaxies of all types and identify the evolutionary paths of individual galaxies through cosmic time. This provides } \added{These reconstructed SFHs can provide } a direct connection between progenitors and descendants, and \deleted{allows } \added{allow } us to constrain when, and how quickly, galaxies formed their stars.

Over the past decade, there have been several algorithms developed to recover SFHs, viz. \textit{MOPED}, \textit{STARLIGHT}, \textit{STECMAP}, \textit{VESPA}, \textit{ULYSS} and \textit{FIREFLY} \citep{heavens2000, fernandes2005, ocvirk2006, tojeiro2007, koleva2009, wilkinson2015}. We develop our own approach in this study to tailor the problem for galaxies at $z\sim1$. The main differences between our algorithm and some of those listed above are the use of composite stellar populations (a group of stars which range in age within a given interval) instead of simple stellar populations (stars born from a single burst in star formation); using a defined set of template spectra which allow for direct comparisons of the SFHs; as well as the assumption of constant star formation within a given time interval. The galaxy spectra are also not continuum-normalised in the fitting process, but photometry is used to calibrate the fluxes.

The \deleted{paper is } \added{goal of this paper is to reconstruct the SFHs of galaxies in the LEGA-C sample and investigate the dependance of individual SFHs on stellar mass, stellar velocity dispersion and star-formation (SF) activity. The paper is } outlined as follows. In Section \ref{sec:data} we give a brief overview of the LEGA-C dataset. In Section \ref{sec:method} we introduce the model for reconstructing the SFHs of the galaxies \deleted{, we present some of the resultant fits, } as well as tests of the model. In Section \added{\ref{fitres}  we present a sample of the resultant fits and general trends of measured parameters. In Section } \ref{sec:mr} we investigate the SFH as a function of stellar velocity dispersion and stellar mass. We demonstrate that we can verify the relation between the evolution of SFHs and mass, and we investigate the \deleted{scatter } \added{variation } in the reconstructed SFHs, at fixed stellar velocity dispersion. Finally, in Section \ref{sec:mr} we \deleted{discuss } \added{summarise } the results. A $\Lambda{CDM}$ model is assumed with $H_0=67.7$\,km\,s$^{-1}$\,Mpc$^{-1}$, $\Omega_m=0.3$ and $\Omega_{\Lambda}=0.7$.

\section{Data}
\label{sec:data}

LEGA-C \citep{vdw2016} is an ongoing ESO Public Spectroscopic survey with VLT/VIMOS of $\sim3000$ galaxies in the COSMOS field ($R.A. = 10^h00^m$; $Dec. = +2^{\circ}12'$). The galaxies were selected from the Ultra-VISTA catalog \citep{muzzin2013f}, with redshifts in the range $0.6<z<1.0$. The galaxies were K-band selected with a magnitude limit ranging from $K(AB) = 21.08$ at z = 0.6 to $K(AB) = 20.7$ at z = 0.8 to $K(AB) = 20.36$ at z = 1.0 (stellar masses $M_* > 10^{10}$\,M$_{\odot}$). These criteria were chosen to reduce the dependence on variations in age, SF activity and extinction, as well as ensure that the  targets were bright enough in the observed wavelength range ($0.6\mu{m}-0.9\mu{m}$) to obtain high quality, high resolution spectra (R\,$\sim3000$). Each galaxy is observed for $\sim20$\,h, which results in spectra with $S/N\sim20$\,\AA$^{-1}$.

The analyses in this work are based on the first-year data release\footnote{http://www.eso.org/qi/catalogQuery/index/93}, which contains spectra of 892 galaxies, 678 of which are in the primary sample and have a $S/N>5$\,\AA$^{-1}$ between rest-frame wavelengths 4000\,{\AA} and 4300\,{\AA} (typically, $S/N\sim20$\,\AA$^{-1}$). Emission line subtracted spectra are used in the fitting algorithm; therefore, the emission line spectrum of each galaxy, computed using the Penalized Pixel-Fitting method \cite[pPXF, ][]{cappellari2004}, is subtracted from the observed spectrum. For details of the emission line fitting procedure, see \deleted{Bezanson et al. (2017)}\added{\citet[][accepted in \apj]{bezanson2018}}. As part of the analysis of the model, we use the following measured quantities: stellar velocity dispersions ($\sigma_*$), 4000\,{\AA} break (D$_n$4000) and H$\delta$ equivalent width \deleted{(EW) } indices [\added{EW(H$\delta$)}], U-V colours, stellar masses ($M_{*,{FAST}}$), UV+IR SFRs, and UV+IR specific SFRs (sSFR$_{UV+IR}$). Stellar masses are determined using FAST \citep{kriek2009} based on photometric measurements, \cite{bruzual2003} stellar population libraries, adopting a \cite{chabrier2003} Initial Mass Function (IMF), \cite{calzetti2000} dust extinction, and exponentially declining SFRs. The UV+IR SFRs are estimated from the UV and IR luminosities, following \cite{whitaker2012}. For details of the data reduction procedure, see \cite{vdw2016}. 

\added{\section{Spectral Fitting Technique}}
\label{sec:method}

\begin{deluxetable}{LRRR}[t]
\tablecaption{Properties of the \textit{FSPS} template spectra.\label{tab:bins}}
\tablewidth{0pt}
\tablehead{
\colhead{Age Bin\tablenotemark{a}} & \colhead{SFR\tablenotemark{b}} & \colhead{M$_*$\tablenotemark{c}} & \colhead{L$_{bol}$\tablenotemark{d}} \\
\colhead{log(yr)} & \colhead{M$_{\odot}$/yr} & \colhead{M$_{\odot}$} & \colhead{log(L$_{\odot}$)}
}
\startdata
0.000-8.000 & 1.000$\times10^{-8}$ & 0.837 & 1.964 \\
8.000-8.300 & 1.005$\times10^{-8}$ & 0.711 & 0.885 \\
8.300-8.475 & 1.010$\times10^{-8}$ & 0.748 & 0.650 \\
8.475-8.650 & 6.750$\times10^{-9}$ & 0.731 & 0.497 \\
8.650-8.750 & 8.646$\times10^{-9}$ & 0.718 & 0.382 \\
8.750-8.875 & 5.332$\times10^{-9}$ & 0.707 & 0.285 \\ 
8.875-9.000 & 3.998$\times10^{-9}$ & 0.695 & 0.187 \\
9.000-9.075 & 5.305$\times10^{-9}$ & 0.685 & 0.127 \\
9.075-9.225 & 2.040$\times10^{-9}$ & 0.671 & 0.099 \\
9.225-9.375 & 1.444$\times10^{-9}$ & 0.652 & -0.043 \\
9.375-9.525 & 1.022$\times10^{-9}$ & 0.639 & -0.161 \\
9.525-9.845 & 2.681$\times10^{-10}$ & 0.618 & -0.347 \\
\enddata
\tablenotetext{a}{Age interval of CSP templates.}
\tablenotetext{b}{SFR s.t. 1\,M$_\odot$ of stars formed within the interval.}
\tablenotetext{c}{Stellar mass (including stellar remnants) with mass loss accounted for.}
\tablenotetext{d}{Bolometric luminosity.}
\end{deluxetable}

\begin{figure}[t]
\centering
\includegraphics[width=0.45\textwidth]{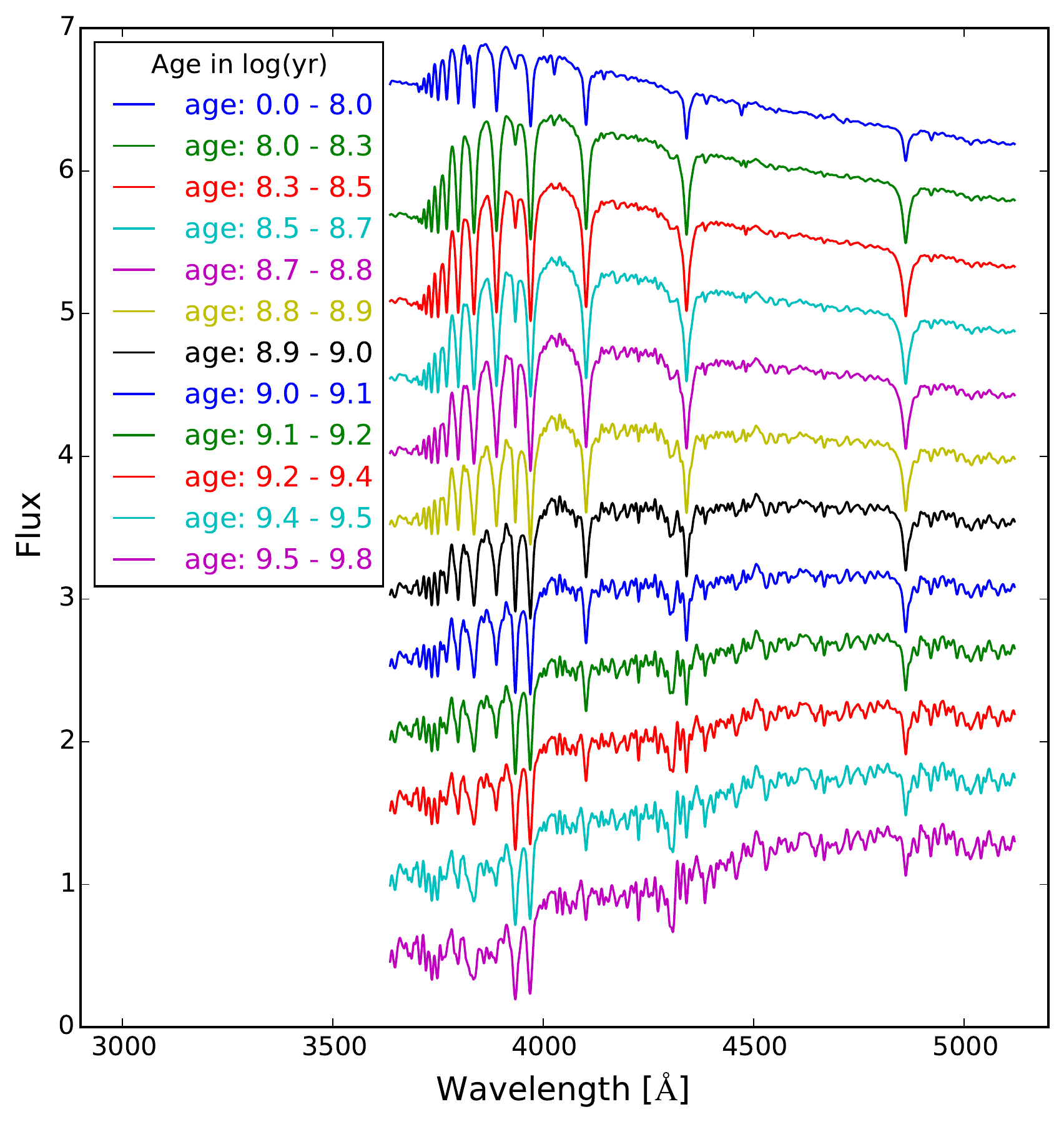}
\caption{Template CSP spectra used to fit LEGA-C galaxies. They were generated from \textit{FSPS}, using the time intervals listed in Table \ref{tab:bins}, with solar metallicity and arbitrary velocity dispersion; and they have been normalised and shifted here for comparison purposes.}
\label{fig:tempbins}
\end{figure}

\added{\subsection{Stellar Population Model}}
\label{sec:model}

To reconstruct the SFHs of galaxies, one needs to gauge the various ages of stellar populations within these galaxies. This is done using stellar population spectra generated with the Python implementation of the Flexible Stellar Population Synthesis package \citep[\textit{FSPS v3.0};][]{conroy2010, conroy2009, foreman2014}, using the MILES spectral library \citep{sanchez2006}, Padova isochrones \citep{girardi2000, marigo2007, marigo2008} and a Kroupa initial mass function \citep{kroupa2001}.

A galaxy spectrum is approximated to be a linear combination of template spectra at varying ages, attenuated by dust:

\begin{equation}
f_{\lambda} = \sum^n_{i=1}{m_i{T_{\lambda,i}10^{-0.4k_{\lambda}E(B-V)_i}}},
\label{eq:a1}
\end{equation}
\begin{displaymath}
{k_{\lambda} = 2.659\Big{(}-2.156 + \frac{1.509}{\lambda} - \frac{0.198}{\lambda^2} + \frac{0.011}{\lambda^3}\Big{)} + 4.05}
\end{displaymath} 

\noindent{where $n$ is the number of stellar population spectra to fit to the galaxy, $T_{\lambda,i}$ are the template spectra, $m_i$ are the weights that scale the templates to match the spectra of the galaxy, $k_\lambda$ is the reddening curve \citep{calzetti2000}, and $E(B-V)_i$ are the dust reddening values.}

We generate 12 composite stellar population spectra (CSPs), with solar metallicity \added{(see Section \ref{sec:robust})}, covering ages from $0$ to about $7$\,Gyrs, the age of the Universe in LEGA-C's redshift range.  To determine the intervals of the 12 age bins of the CSPs, simple stellar population spectra (SSPs) were generated and the cumulative absolute difference from one spectrum to another was calculated as the age was increased; which was then divided into 12 percentiles with equal width (see Table \ref{tab:bins} and Figure \ref{fig:tempbins} for the properties of the CSPs in each age bin). This method of determining the time intervals generates template spectra that optimise the temporal sampling of an evolving stellar population. In practice, the age bins are $\sim0.15$\,dex wide over the age range $0-7$\,Gyrs.

The template spectra are generated with a constant SFR and are normalised such that $1$\,M$_{\odot}$ of stars are formed within each time bin (stellar masses include stellar remnants). Note that there is mass loss in each bin as massive stars die off (see Table \ref{tab:bins}). \added{We assume a constant SFR within each time bin as it presents a more realistic evolution of a galaxy's star formation with time, and can take into account rapid changes in the SFH. Choosing SSP templates would not lead to significantly different SFHs, however, it would lead to aliasing effects when reconstructing the SFHs for samples of galaxies. } The templates are also broadened to the velocity dispersion of the galaxies \deleted{. } \added{\cite[][accepted in \apj]{bezanson2018}.} It is assumed that dust reddening is the same for all populations except for the youngest population (age $<100$\,Myrs). Dust extinction is expected to be different for young stellar populations as they are usually observed to be nested in the dust of their molecular birth clouds \citep{charlot2000}. Therefore, two dust reddening values are fit for: $E(B-V)_{1}$, for the age range $0-100$\,Myrs, and $E(B-V)_{2}$, for the rest of the age ranges.

\deleted{Solar metallicity was used to generate all the CSPs because according to 
\cite{gallazzi2005, gallazzi2014}
, the metallicity-mass relation flattens out to solar metallicity in LEGA-C's mass range (log(M)\,$\gtrsim 10.5$), for $z\sim0.7$ galaxies. We also found that using sub-solar or super-solar metallicities for the templates, instead of solar metallicity, generally results in no significant differences in the $\chi^2$ values of the fits. Nevertheless, if we assign the implausibly high ($2.5$\,Z$_\odot$) or low ($0.4$\,Z$_\odot$) metallicities for the galaxies in our sample, the derived ages do not systematically change by more than $0.1$\,dex (less than the width of one age bin).
}

\subsection{Fitting Algorithm}
\label{sec:fit_alg}

To find the optimal values for the 14 parameters, viz. the 12 weight factors ($m_i$) \added{for the 12 CSP templates } and 2 dust reddening values ($E(B-V)_i$), we used \textit{emcee}, a Python implementation of an affine invariant ensemble sampler for MCMC \citep{foreman2013} which was proposed by \cite{goodman2010}. It uses MCMC `walkers' which randomly explore the parameter space, where each proposed step for a given walker depends on the positions of all the other walkers in the ensemble, with the aim of converging to the most likely parameter values.

The priors for the 14 parameters were set such that all parameter values were always greater or equal to 0, and the upper limit for $E(B-V)_i$ was set to 3. The parameter value for the youngest bin was initially set to be equal to the measured SFR from UV+IR measurements, but it was allowed to vary between 1/3 and 3 times that value during the fitting process, allowing for measurement errors. For the other bins, the best fitting single template, computed using least-squares fitting, was assigned all the stellar mass, with all other parameter values set to $10^{-6}$. Starting with equal SFRs in all bins also recovers the SFHs, however, the algorithm may take longer to converge to the optimal values.

For each galaxy, 100 MCMC walkers were used, initiated in a small region around the starting values mentioned above. A total of $20000$ samples were taken and 1000 steps were kept after burn-in. The mean acceptance fraction was $\gtrsim0.2$ and the typical autocorrelation time was $\sim95$ iterations. The optimal values for the parameters are taken as the 50$^{th}$ percentile of the list of samples of the converged walkers, and the lower and upper uncertainties are the 16$^{th}$ and 84$^{th}$ percentiles, respectively. The fitting algorithm resulted in 607 good fits based on their normalised $\chi^2$ values ($<5$, from visual inspection of the fits)\deleted{. 
}\added{, and these were used in the analyses. The spectra that were not well-fit were mainly due to low S/N and AGN.}

\begin{figure*}[t]
\gridline{\fig{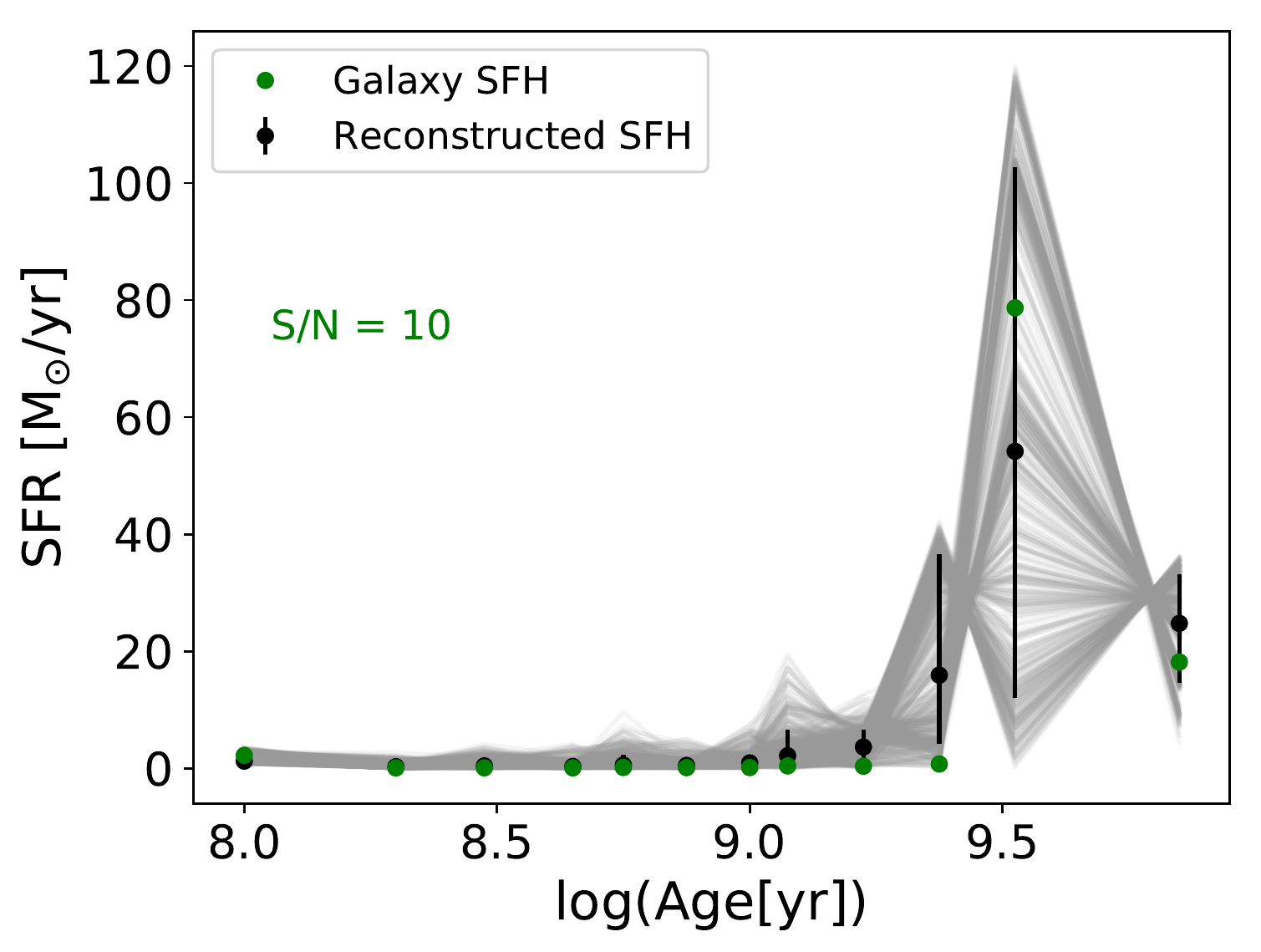}{0.45\textwidth}{}
\fig{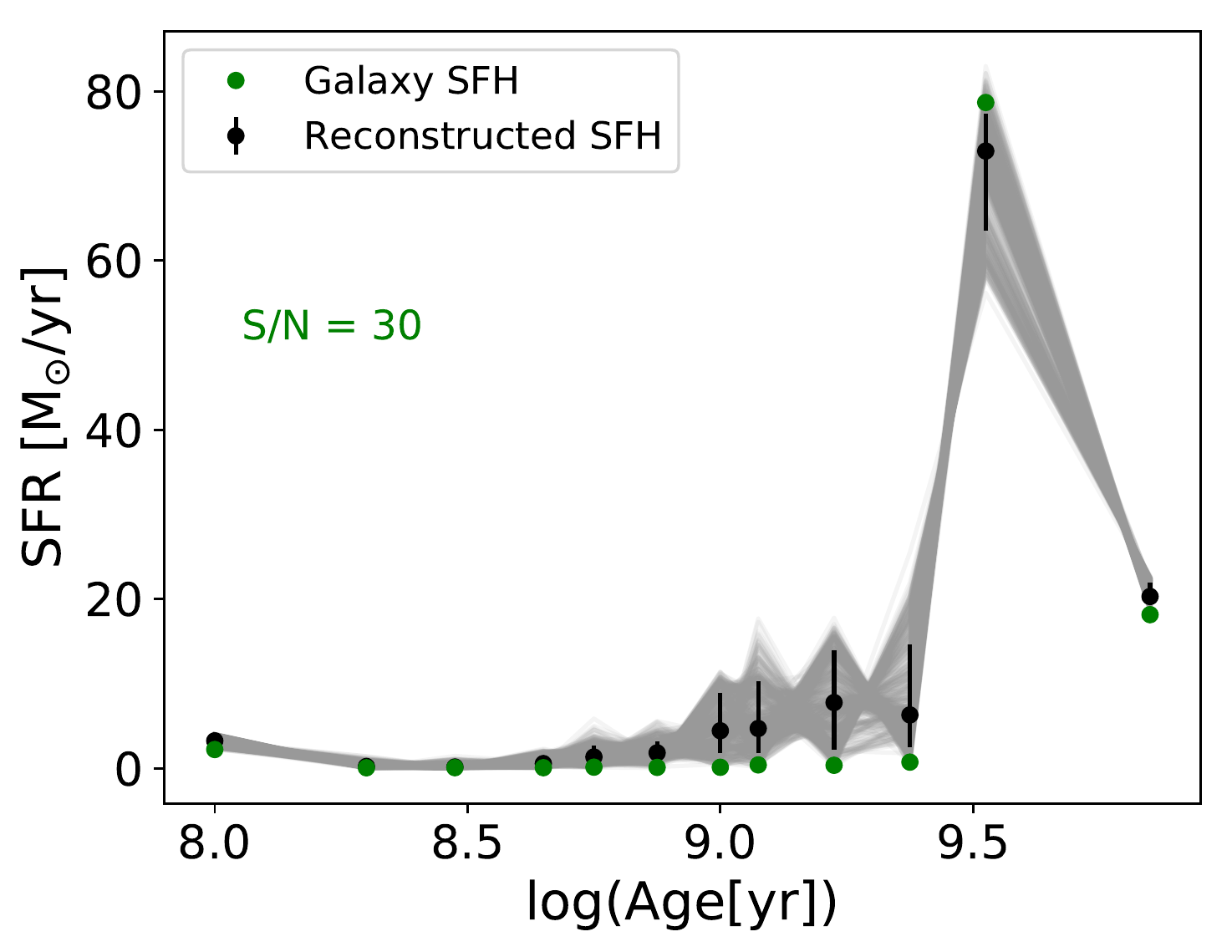}{0.45\textwidth}{}}
\caption{\added{Reconstructed SFH (black) of a synthetic galaxy (green) with S/N\,$=10$\,\AA$^{-1}$ (left) and S/N\,$=30$\,\AA$^{-1}$ (right). The converged walkers are shown in grey and the upper and lower uncertainties are based on the 16$^{th}$ and 84$^{th}$ percentiles, respectively, as explained in Section \ref{sec:fit_alg}. By S/N\,$=30$\,\AA$^{-1}$, the recovered SFHs predict the stellar mass, age and luminosity with precision $\leq0.1$\,dex.}}
\label{fig:synsfhs}
\end{figure*}

\added{\subsection{Robustness of Fitting Results}}
\label{sec:robust}

\added{To assess  the robustness of the model, we performed the following tests: generate and fit synthetic spectra; compare model stellar mass measurements of the LEGA-C population with those obtained from broad-band photometry (see Section \ref{sec:data}); fit a sample of SDSS spectra and compare model stellar masses with literature measurements; and test the assumption of solar metallicity.} 

\added{Synthetic galaxy spectra were generated with varying SFHs using the CSPs in Section \ref{sec:model}, including simulated noise that mimics LEGA-C variance spectra, to compare how well the algorithm recovered the SFHs. 20 SFHs were generated for each S/N (5, 10, 20, 30, 40, 50 and 60\,\AA$^{-1}$), and the average deviations of the true \amw, stellar mass and luminosity from the best-fitting model parameters were computed. In general, the model sufficiently recovered the SFHs, however, we note that the quality of the results depends on the noise introduced into a spectrum (see Figure \ref{fig:synsfhs} for two examples). Stellar mass and luminosity are recovered with precision $\leq0.1$\,dex for S/N$\geq20$, while \amw\, only requires S/N$\geq10$ to reach the same level of precision. We note that these tests only constrain the purely random uncertainties due to the noise in the spectra while they do not include systematic errors in the data (e.g., sky subtraction, flux calibration) and systematic uncertainties in the FSPS model spectra.} 

\added{Imposing the MCMC model on the LEGA-C dataset and comparing the stellar masses measured from the model to those measured from FAST (using photometric measurements), resulted in very good agreement between the two methods, with a scatter of $\sim0.2$\,dex and an offset of $\sim0.03$\,dex. This scatter is larger than the formal uncertainty on our mass measurements ($\sim0.15$\,dex).} 

\added{We used the fitting algorithm on a sample of 20 SDSS spectra of massive local galaxies ($z\sim0.1$), selected by stellar mass (M\,$_*>10^{10}$), to determine whether the model could recover the stellar masses measured in the literature. We compared the model stellar masses to measurements from the Portsmouth method \citep{maraston2009} and found satisfactory agreement, with a $\sim0.2$\,dex scatter. The maximum age of the templates was increased to $\sim12$\,Gyrs to account for the low redshift ($\sim0.1$) of the SDSS galaxies. The $0.2$\,dex random uncertainty is an indication of how results vary as a consequence of using a different SPS model (here, \cite{maraston2009} vs. }\textit{\added{FSPS}}\added{) and fitting algorithm. } 

\added{Solar metallicity was used to generate all the CSPs because according to 
\cite{gallazzi2005, gallazzi2014}, the metallicity-mass relation flattens out to solar metallicity in LEGA-C's mass range (log(M)\,$\gtrsim 10.5$), for $z\sim0.7$ galaxies. On the other hand, \cite{jorgensen2017} find evidence for evolution in the metallicity for cluster galaxies, as well as a trend of increasing metallicity with increasing velocity dispersion. We test our approach by repeating our fits with implausibly low metallicity ($0.4$\,Z$_\odot$, sub-solar) and high metallicity ($2.5$\,Z$_\odot$, super-solar) CSPs. We find no significant differences in the $\chi^2$ values of the fits, but, naturally, the inferred ages depend on the chosen metallicity. If we assign sub-solar metallicity for galaxies in our sample, the derived mass-weighted and light-weighted ages are older by $0.05$ and $0.08$\,dex, respectively, with a standard deviation of $0.16$ and $0.24$\,dex, respectively. If we assign super-solar metallicity for the sample, the light-weighted ages are younger by $0.03$\,dex, with a standard deviation of $0.20$\,dex. The mass-weighted age changes from solar to high metallicity are not normally distributed: 80\% of the galaxies have the same age to within $0.20$\,dex, while for the remaining 20\% the change in age ranges from $0.2$ to $0.9$\,dex. However, only 10 of these galaxies' mass-weighted ages change by $\geq$\,$0.5$\,dex and they have a mean light-weighted age of $\sim0.4$\,Gyr. The age changes do not depend on the measured stellar mass or stellar velocity dispersion.} 

\added{The velocity dispersion-metallicity trend presented by \citet{jorgensen2017} implies that our assumption of solar metallicity for all galaxies may introduce a correlation between velocity dispersion and age. Our tests show that across the velocity dispersion range \sigs\,$=100-250$\,km\,s$^{-1}$, the magnitude of this effect would be at most $0.15$\,dex and likely less. This potential bias is insufficient to explain the \sigs-age relation we find in Section \ref{sec:insfh}. Follow-up studies that explore the interdependence of age, metallicity and other galaxy properties will need to take metallicity variations into account.}

\begin{figure*}[t]
\gridline{\fig{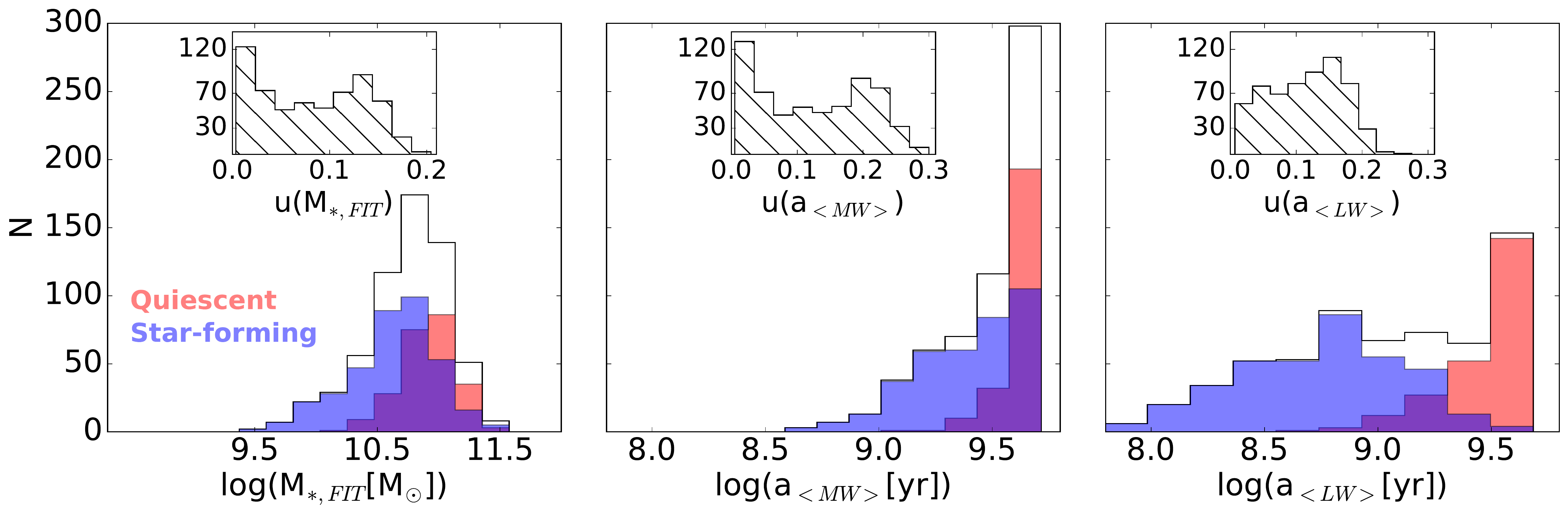}{0.95\textwidth}{}}
\caption{Distributions of \mfit\,(left), \amw\,(middle) and \alw\,(right) of the LEGA-C sample. The quiescent and star-forming populations (as defined in Section \ref{sec:corr}) are shown in red and blue, respectively. The distribution of the uncertainties for each parameter are shown at the top of each figure.}
\label{fig:hist}
\end{figure*}

\added{\section{Fitting Results}}
\label{fitres}

\subsection{Model Outputs}

Figure \ref{fig:hist} shows the distribution of the model-measured stellar masses (\mfit, left panel), mean mass-weighted ages\footnote{Mean mass-weighted and light-weighted ages are obtained by averaging the midpoint ages of the CSPs weighted by luminosity and mass,  respectively.\label{n:amlw}} (\amw, middle panel) and mean light-weighted ages\textsuperscript{\ref{n:amlw}} (\alw, right panel) of the LEGA-C sample, along with the distribution of uncertainties for each parameter. The distributions are separated into the quiescent (red) and star-forming (blue) populations to show the differences in the distributions based on current SF activity (see Section \ref{sec:mr}). The galaxies in the LEGA-C sample span a broad range of ages: \alw\, can be as  young as $60$\,Myrs and as old as $4.8$\,Gyrs, and has a median value of $1.2$\,Gyrs (see Figure \ref{fig:hist}). \amw\, ranges from about $400$\,Myrs to about $5.2$\,Gyrs, with a median value of $3.8$\,Gyrs. However, most of these galaxies are old, with about 60\% being older than $3$\,Gyrs. The \mfit\, of the galaxies ranges from $\sim2\times10^{9}$\,M$_{\odot}$ to $\sim4\times10^{11}$\,M$_{\odot}$, with a median value of about $6\times10^{10}$\,M$_{\odot}$. The \added{formal } age and mass uncertainties lie in the ranges 1-60\% and 1-40\%, respectively. \deleted{We note that these uncertainties are underestimated as they are computed from the MCMC model fit and they } \added{As stated in Section \ref{sec:robust}, these uncertainties } do not include systematic errors\deleted{on the data as well as the template spectra (stellar population synthesis model)}.

\addtocounter{subsubsection}{-1}
\begin{figure*}
\gridline{\fig{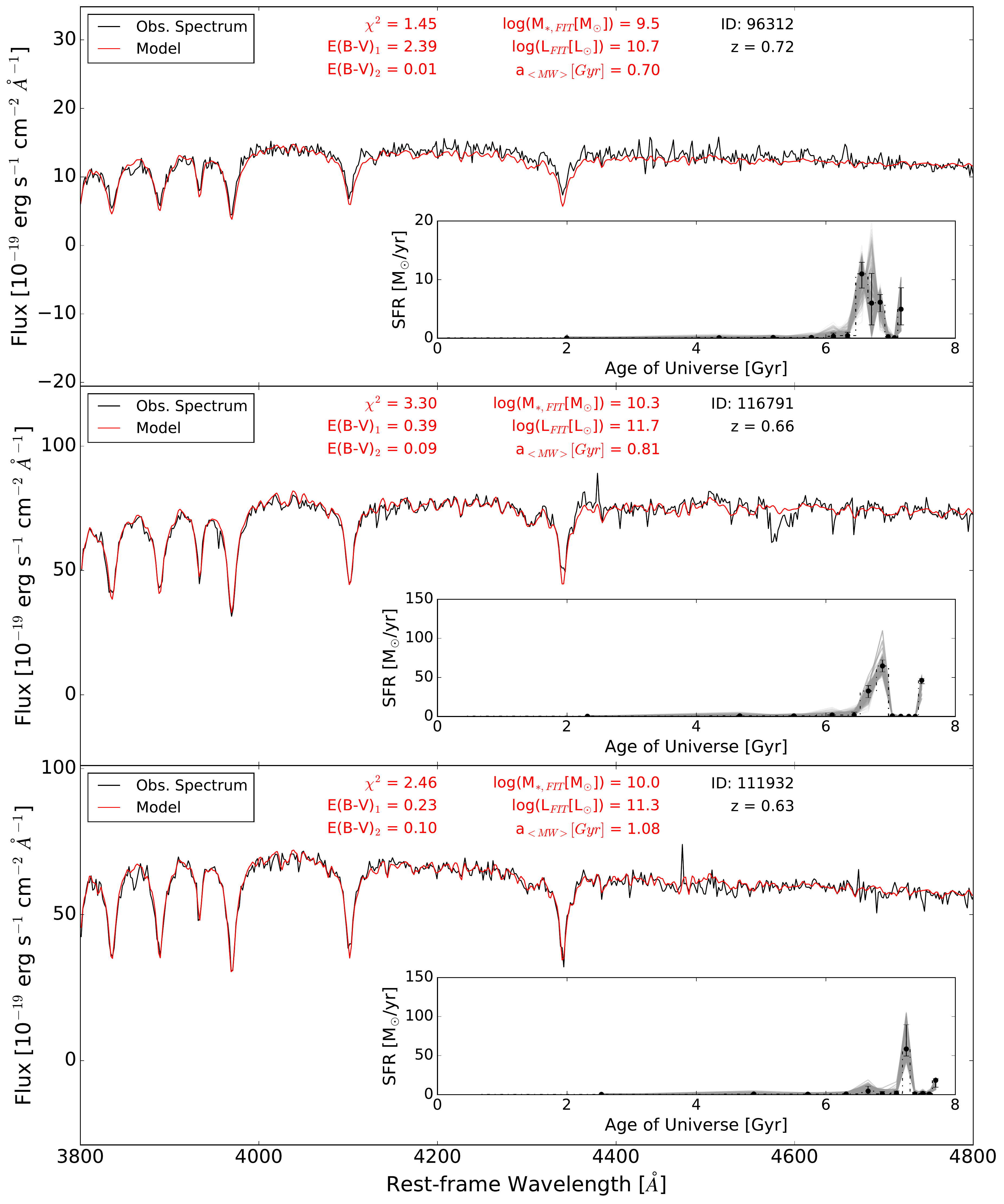}{0.95\textwidth}{}}
\caption{\added{Sample of emission line subtracted spectra of 12 LEGA-C galaxies with the best fitting model obtained from combining the 12 template spectra using MCMC. The bottom-right figure, in each plot, is the reconstructed star formation history (the converged walkers are shown in grey). The MCMC resultant mass, luminosity, mass-weighted age and dust reddening values are shown in red. The spectra are ordered by \amw.}}
\label{fig:fit1}
\end{figure*}

\renewcommand{\thefigure}{\arabic{figure} (Continued)}
\addtocounter{figure}{-1}

\begin{figure*}
\gridline{\fig{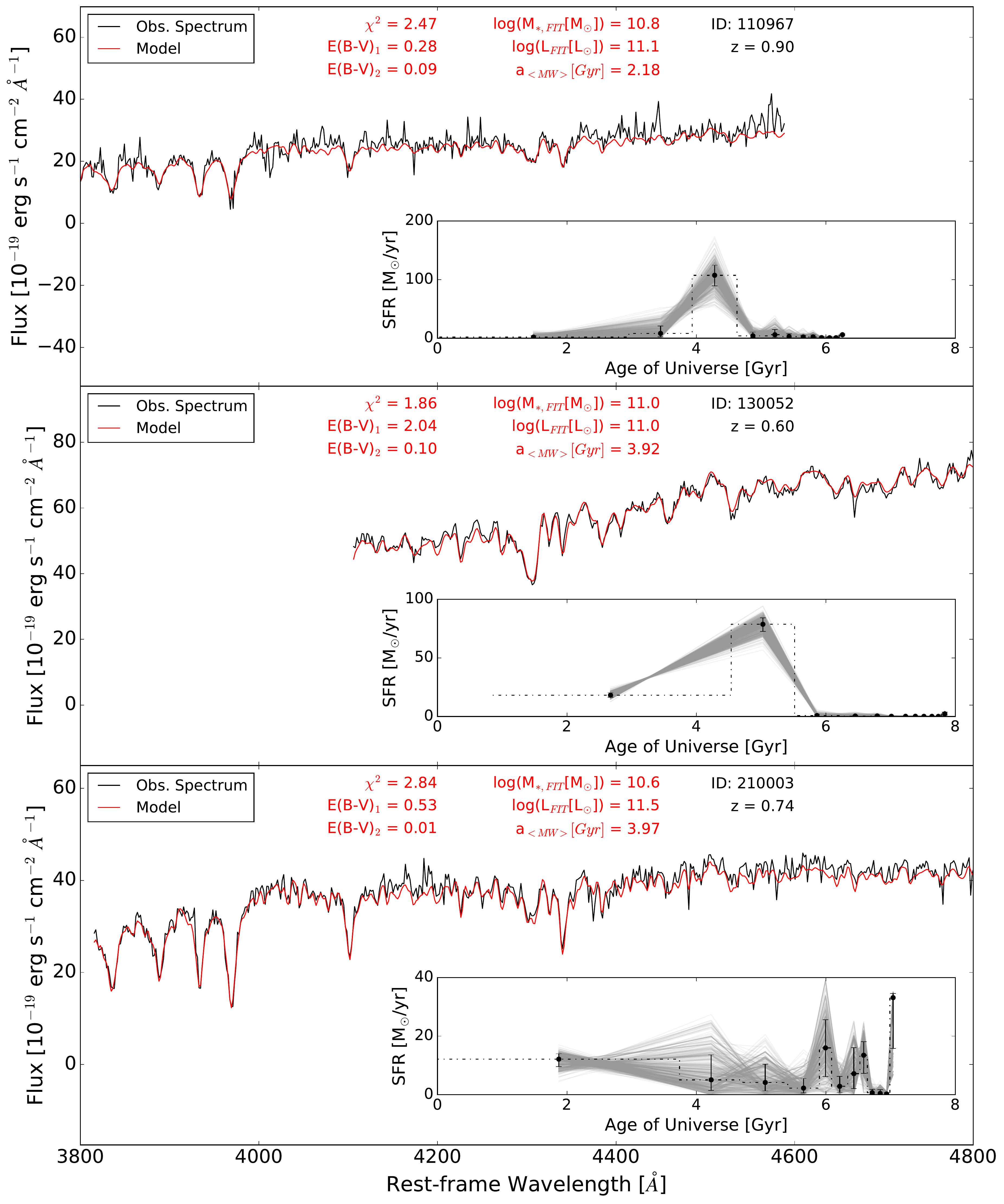}{0.95\textwidth}{}}
\caption{}
\label{fig:fit2}
\end{figure*}

\addtocounter{figure}{-1}

\begin{figure*}
\gridline{\fig{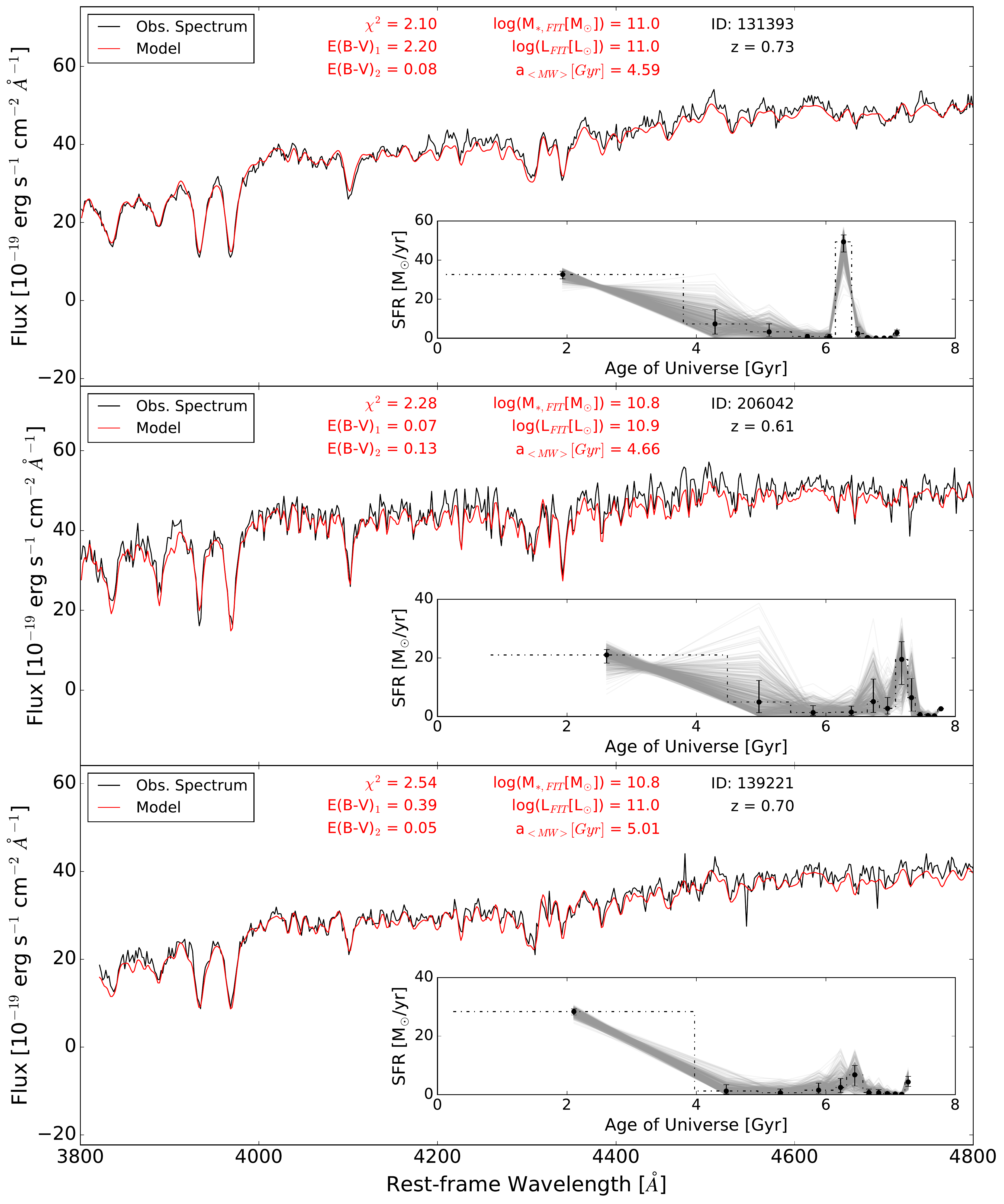}{0.95\textwidth}{}}
\caption{}
\label{fig:fit3}
\end{figure*}

\addtocounter{figure}{-1}

\begin{figure*}[t]
\gridline{\fig{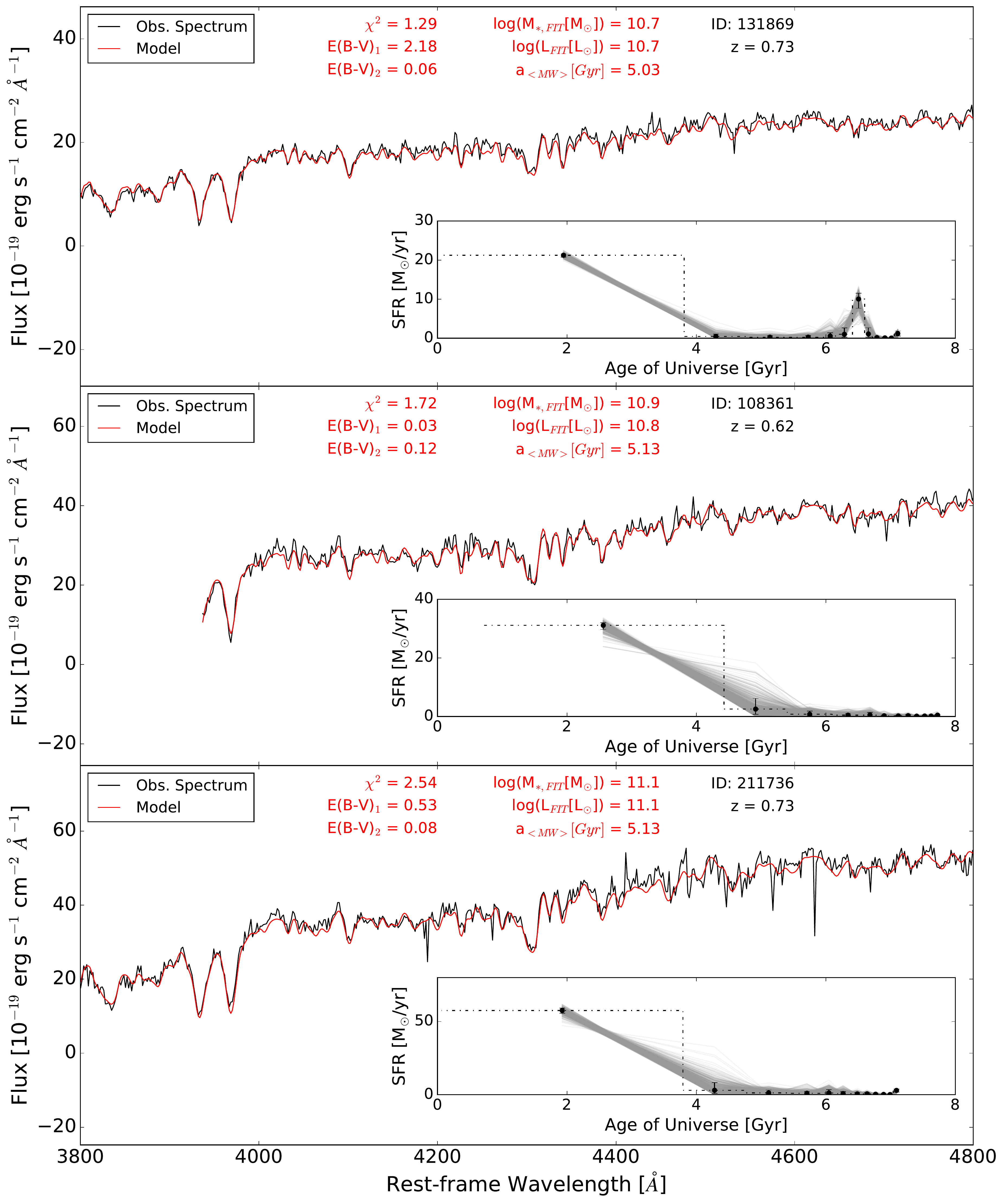}{0.95\textwidth}{}}
\caption{}
\label{fig:fit4}
\end{figure*}

\renewcommand{\thefigure}{\arabic{figure}} 

\begin{figure*}[t]
\gridline{\fig{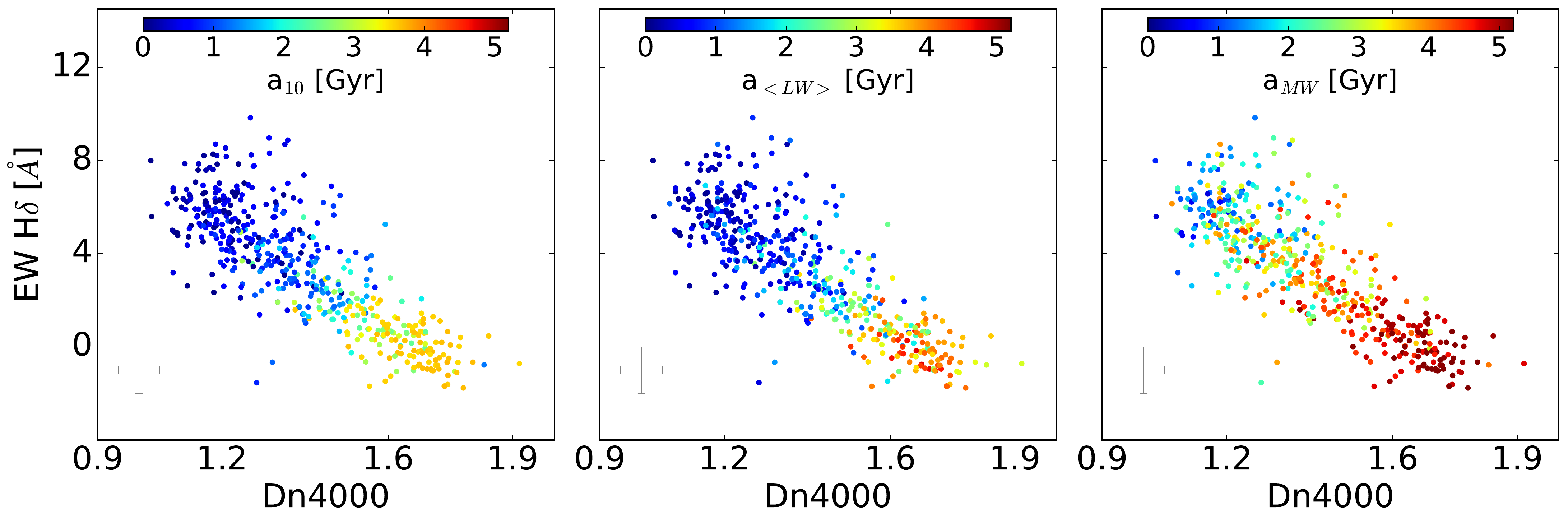}{0.9\textwidth}{(a)}}
\label{fig:hdd4000_a}
\gridline{\fig{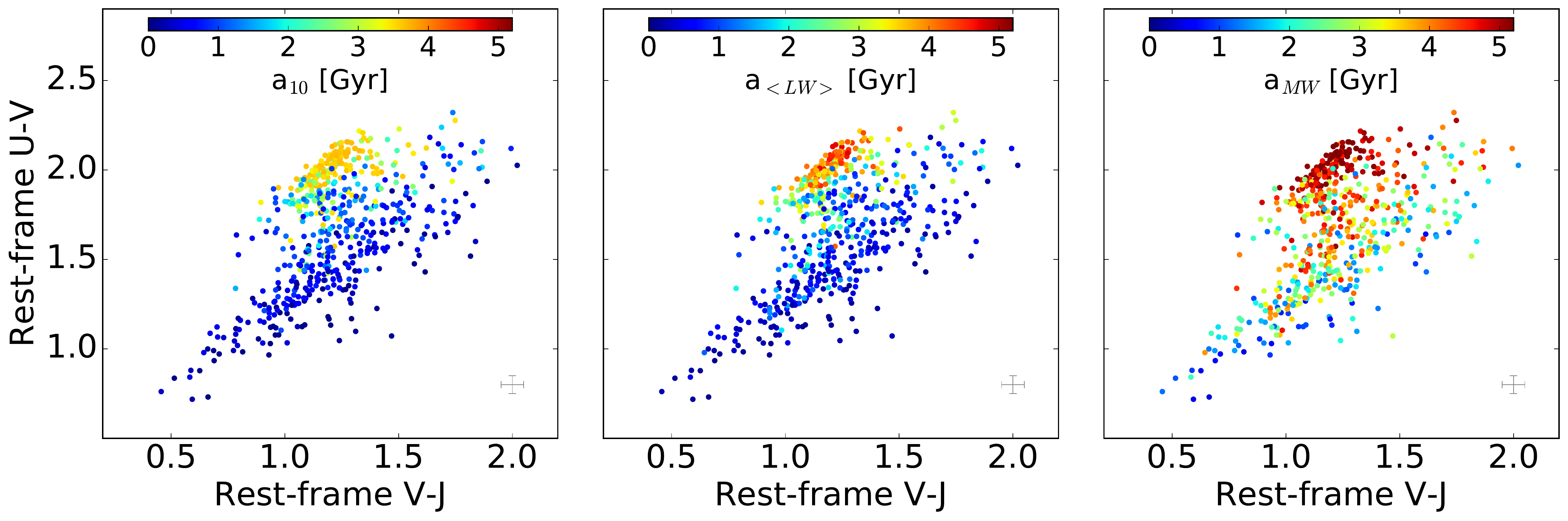}{0.9\textwidth}{(b)}}
\label{fig:hdd4000_b}
\caption{\added{EW(H$\delta$) versus D$_n$4000 (upper panel) and U-V colour versus V-J colour (lower panel) colour-coded by the time after which the final 10\% of stars were formed (left), the mean light-weighted age (middle), and the mean mass-weighted age (right). Typical error bars are indicated in grey.}}
\label{fig:hdd4000}
\end{figure*}

\added{\subsection{Sample SFHs}}
\label{sec:insfh}

Figure \ref{fig:fit1} shows the spectra of a sample of LEGA-C galaxies \added{(in \amw\, order) } along with the best-fitting model spectra as described by Equation \ref{eq:a1} using the optimal parameter values from \textit{emcee}. The weight factors, $m_i$, represent the star formation histories of these galaxies (shown on the bottom-right of each figure). The resultant normalised $\chi^2$, dust reddening values (\ebv), stellar masses (\mfit), luminosities (\lfit) and mean mass-weighted ages (\amw) from the model are shown in red. The sample was selected to display the wide range of SFHs recovered.

The reconstructed SFHs reveal that although most galaxies at $z\sim1$ have \amw\,$>3$\,Gyrs, the sample \deleted{has a good variety } \added{spans a wide range } of histories. For the older massive galaxies, the oldest template (stars in the age range $\sim$3-7\,Gyrs) contributes to the majority of their mass. Some of these galaxies only contain the oldest stars and have since been quiescent, i.e. for the past $\sim3$\,Gyrs (see the SFHs of 108361, 211736 and 130052 in Figure \ref{fig:fit1}). However, some galaxies were quiescent for several Gyrs and then had a renewed period of growth, either due to SF rejuvenation, or merging with a younger population. A merger could result in either an integration of the younger population with no further activity, or trigger bursts of star formation. This young population of stars accounts for $\sim10$\,\% of the mass of these galaxies (e.g. 206042, 131869 and 131393 in Figure \ref{fig:fit1}). We will explore the frequency of such rejuvenation events in more detail in a follow-up study.

\addtocounter{subsubsection}{-1}
\added{\subsection{General Trends}}
\label{sec:trends}

{
\deleted{Comparison between total stellar masses measured by the fitting algorithm ($M_{*,{FIT}}$) to those obtained by FAST based on photometric measurements ($M_{*,{FAST}}$). The MCMC mass and its upper and lower uncertainties are based on the 50$^{th}$, 16$^{th}$ and 84$^{th}$ percentiles, respectively, as explained in Section \ref{sec:fit_alg}. }}


{
\deleted{H$\delta$ EW versus D$_n$4000 (upper panel) and U-V colour versus V-J colour (lower panel) colour-coded by the time after which the final 10\% of stars were formed (left), the mean light-weighted age (middle), and the median stellar age (right). Typical error bars are indicated in grey.}}

\deleted{To assess  the robustness of the model, we performed the following tests: generate and fit synthetic spectra; fit a sample of SDSS spectra and compare model stellar masses with literature measurements; compare model stellar mass measurements of the LEGA-C population with those obtained from FAST; and analyse trends of familiar spectral features and colours with model masses and ages.
 }

\deleted{Synthetic galaxy spectra were generated with varying SFHs, including simulated noise that mimics LEGA-C variance spectra, to compare how well the algorithm recovered the SFHs. 20 SFHs were generated for each S/N, and the average deviations of the true \amw, stellar mass and luminosity from the best-fitting model parameters were computed. In general, the model sufficiently recovered the SFHs, however, we note that the quality of the results depends on the noise introduced into a spectrum. Stellar mass and luminosity are recovered with precision $\leq0.1$\,dex for S/N$\geq20$, while \amw\, only requires S/N$\geq10$ to reach the same level of precision. As stated in Section \ref{sec:insfh}, the uncertainties computed here are underestimated as they are only based on the fitting algorithm.
}

\deleted{We used the fitting algorithm on a sample of 20 SDSS spectra of massive local galaxies ($z\sim0.1$), selected by stellar mass (M\,$_*>10^{10}$), to determine whether the model could recover the stellar masses measured in the literature. We compared the model stellar masses to measurements from the Portsmouth method 
\citep{maraston2009} 
and found satisfactory agreement, with a $\sim0.2$\,dex scatter. The maximum age of the templates was increased to $\sim12$\,Gyrs to account for the low redshift ($\sim0.1$) of the SDSS galaxies. The $0.2$\,dex random uncertainty is an indication of how results vary as a consequence of using a different SPS model (here, 
\cite{maraston2009} 
vs. }\textit{\deleted{FSPS}}
\deleted{). Imposing the MCMC model on the LEGA-C dataset and comparing the stellar masses measured from the model to those measured from FAST (using photometric measurements), resulted in very good agreement between the two methods, with a scatter of $\sim0.2$\,dex and an offset of $\sim0.03$\,dex (See Figure \ref{fig:masscomp}). This scatter is larger than the formal uncertainty on our mass measurements ($\sim0.15$\,dex).
}


\deleted{The upper panel of} \added{Figure \ref{fig:hdd4000}(a)} presents the distribution of \deleted{the }\added{EW(}H$\delta$\deleted{EW}\added{)} as a function of the D$_n$4000 break colour-coded by the time after which the final 10\% of stars were formed  (\aft, left panel), \deleted{\alw } {\added{\alw}} (middle panel) \deleted{, and the median stellar age}\footnote{\deleted{Median ages are obtained by computing the time after which 50\% of the stellar population was formed.}} 
\addtocounter{footnote}{-1}
\deleted{(\amed, } \added{and }{\added{\amw}} \added{(}right panel), estimated from the model. \added{The EW(H$\delta$)-D$_n$4000 distribution is analysed in depth in 
\cite{wu2018}. } As expected, for all three age parameters, galaxies generally evolve from the upper-left region (high EW(H$\delta$) and low D$_n$4000) to the lower-right region (low EW(H$\delta$) and high D$_n$4000) as they age. \deleted{\aft and \alw } {\added{\aft}} \added{and }{\added{\alw}} are more correlated with each other than \deleted{\amed } {\added{\amw}} because they track young stars; they also have smoother transitions in the EW(H$\delta$)-D$_n$4000 plane because those features primarily track recent SF activity ($\lesssim1$\,Gyr). \deleted{The lower panel of} \added{Figure \ref{fig:hdd4000}(b)} shows the rest-frame U-V colour as a function of restframe V-J colour-coded by the same 3 age parameters as above. Once again, expected trends are seen: \deleted{\aft, \alw and \amed } {\added{\aft}}\added{, }{\added{\alw}} \added{and }{\added{\amw}} correlate with the restframe colours as U-V and V-J primarily reflect recent star formation ($\sim1$\,Gyr). There is a notable population of old galaxies (\deleted{\amed}\added{\amw} \,$>3.5$\,Gyrs) with relatively blue colours, which indicates that these galaxies have extended SFHs. 

To demonstrate the validity of the old galaxies (\deleted{\amed,}\added{\amw\,} $> 3.5$\,Gyrs) in the  young region of the EW(H$\delta$)-D$_n$4000 plane, i.e. galaxies in red in Figure \ref{fig:hdd4000}'s right panel, with D$_n$4000 $<1.3$ and EW(H$\delta$) $>2$, we refer to their SFHs. These galaxies formed most of their stars early on, but also have significant recent star formation. While some seem to have been quiescent at some point in their history before they were possibly rejuvenated or merged with another galaxy (e.g. 206042 in Figure \ref{fig:fit1}), others formed stars throughout their history (e.g. 210003 in Figure \ref{fig:fit1}). Moreover, the presence of young and old populations can be seen in their spectra: they have clearly visible Balmer lines, characteristic of younger galaxies; but they also have H and K absorption lines of singly ionized Calcium with similar strengths, which is typical of older galaxies, in addition to the presence of the G-band (absorption lines of the CH molecule) around 4300\,{\AA}. As a test, we reran the fits of these galaxies excluding the 3 oldest templates and found that the spectra cannot be well fit.

{
\deleted{Velocity dispersion versus the model-measured stellar mass colour-coded by \amw. The star-forming and quiescent populations are shown in the middle and right panels, respectively. Typical error bars are indicated in grey. The clear separation between young and old galaxies at $\sigma_*\sim170$\,km\,s$^{-1}$ shows a stronger correlation between \amw\, and }
\deleted{\, over \mfit, which also depends on the current SF activity.}}

There is also a population of galaxies that seem to contain only young stars (e.g. 111932 and 116791 in Figure \ref{fig:fit1}), which would imply that these galaxies formed more than 90\% of their mass recently (\deleted{lookback}\added{when the Universe was $>6$} \,\deleted{$<8$\,} Gyrs). To test if young populations are `outshining' the rest of these galaxies, i.e. if there are hidden populations of old stars, we reran the fits of these galaxies allowing only the oldest template parameter to vary. We found that the contribution in mass of the old population can increase by $\sim5-10$\% before the \deleted{fits are visually degraded} \added{normalised $\chi^2$ changes by more than $0.08$\,dex. The change in $\chi^2$ is mainly due to the continuum shape of the spectra. Therefore, } \deleted{, therefore,} these galaxies do not harbour significant populations of old stars that are concealed by the light of very young stars.


\begin{figure*}[t]
\gridline{\fig{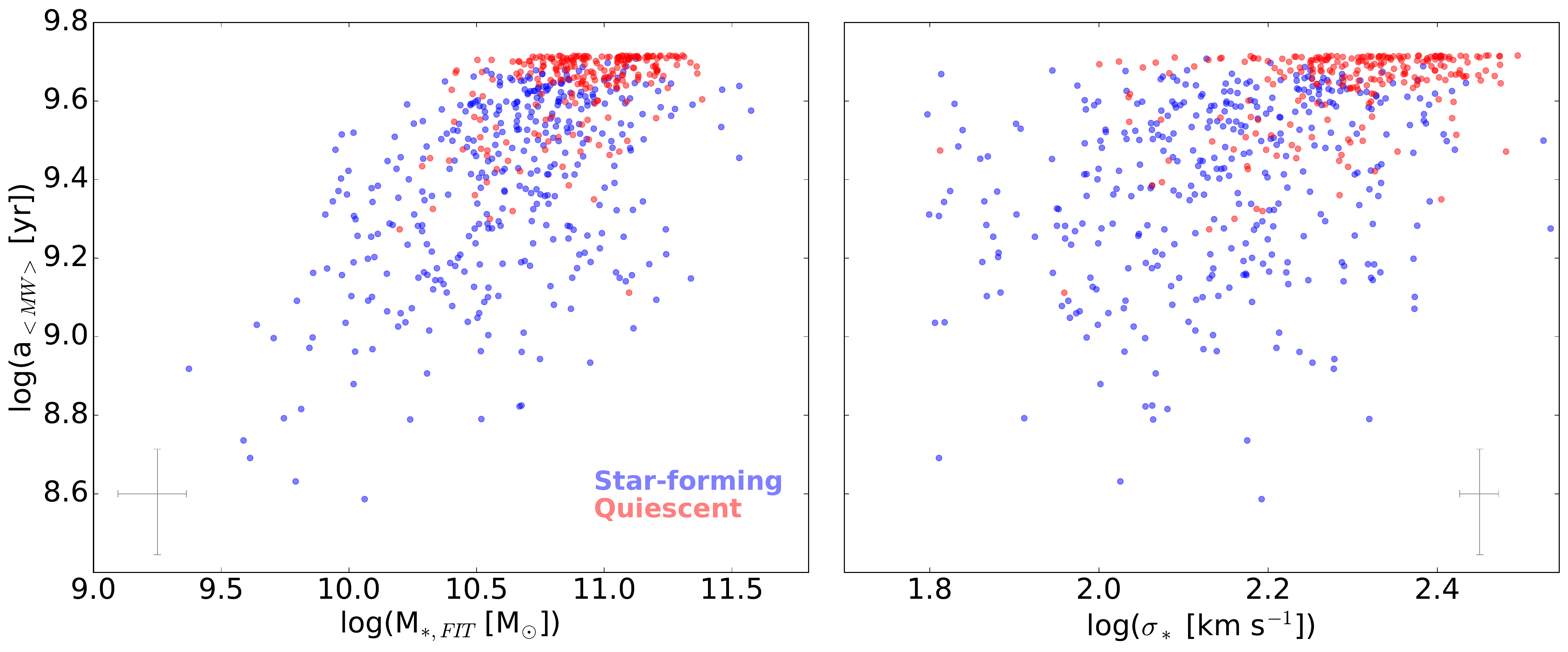}{0.85\textwidth}{}}
\caption{\added{\amw\, as a function of \mfit\, (left) and \sigs\, (right). The star-forming and quiescent populations are indicated in blue and red, respectively, and typical error bars are indicated in grey. Galaxies with \sigs\,$\gtrsim200$\,km\,s$^{-1}$ are almost exclusively old ($>4$Gyrs) and quiescent, which indicates that \sigs\, is a stronger predictor of age and SF activity.}}
\label{fig:age_vs_v_m}
\end{figure*}

\begin{figure*}[t]
\gridline{\fig{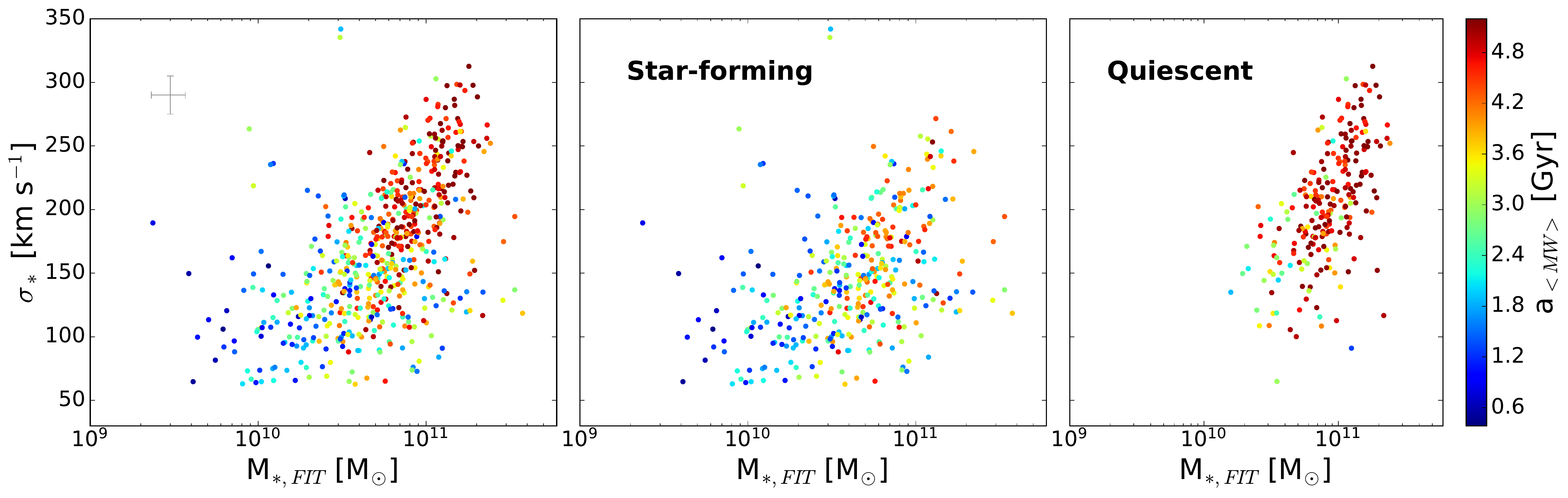}{1.0\textwidth}{}}
\caption{\sigs\, versus \mfit, colour-coded by \amw. The star-forming and quiescent populations are shown in the middle and right panels, respectively. Typical error bars are indicated in grey. The clear separation between young and old galaxies at $\sigma_*\sim170$\,km\,s$^{-1}$ shows a stronger correlation between \amw\, and \sigs\, over \mfit, which also depends on the current SF activity.}
\label{fig:vdisp_vs_mass}
\end{figure*}

\begin{figure*}[t]
\gridline{\fig{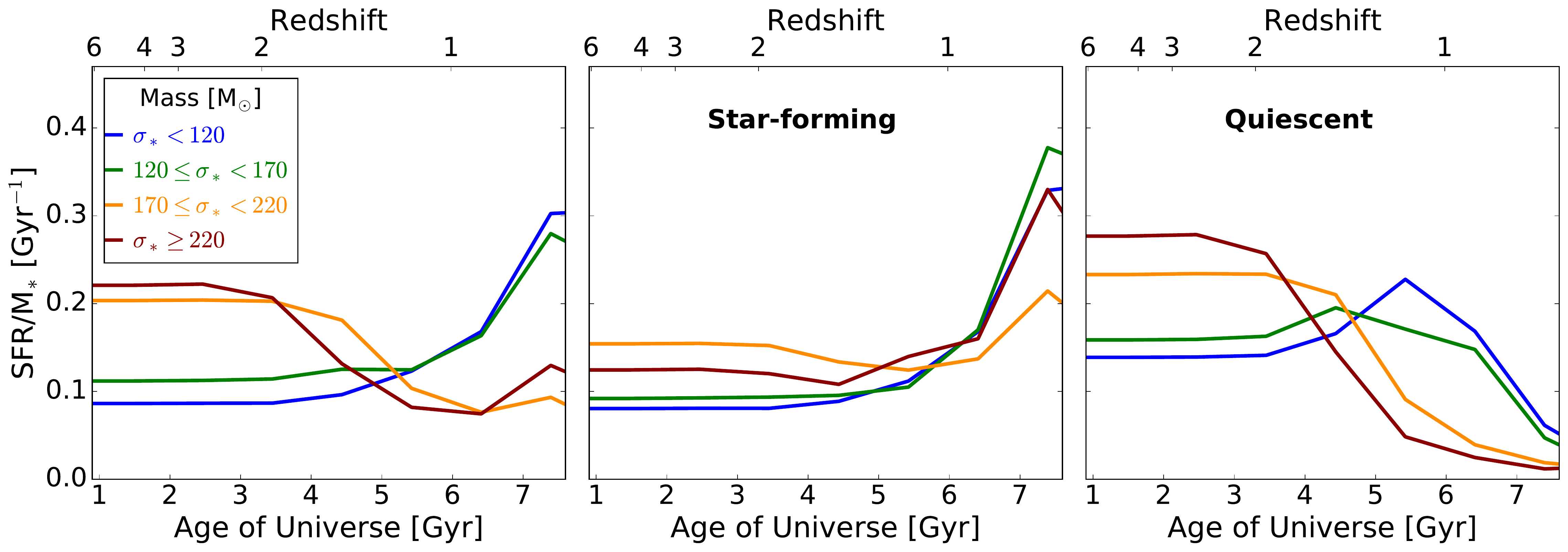}{0.95\textwidth}{(a)}}
\label{fig:sfr_lb_mbins_a}
\gridline{\fig{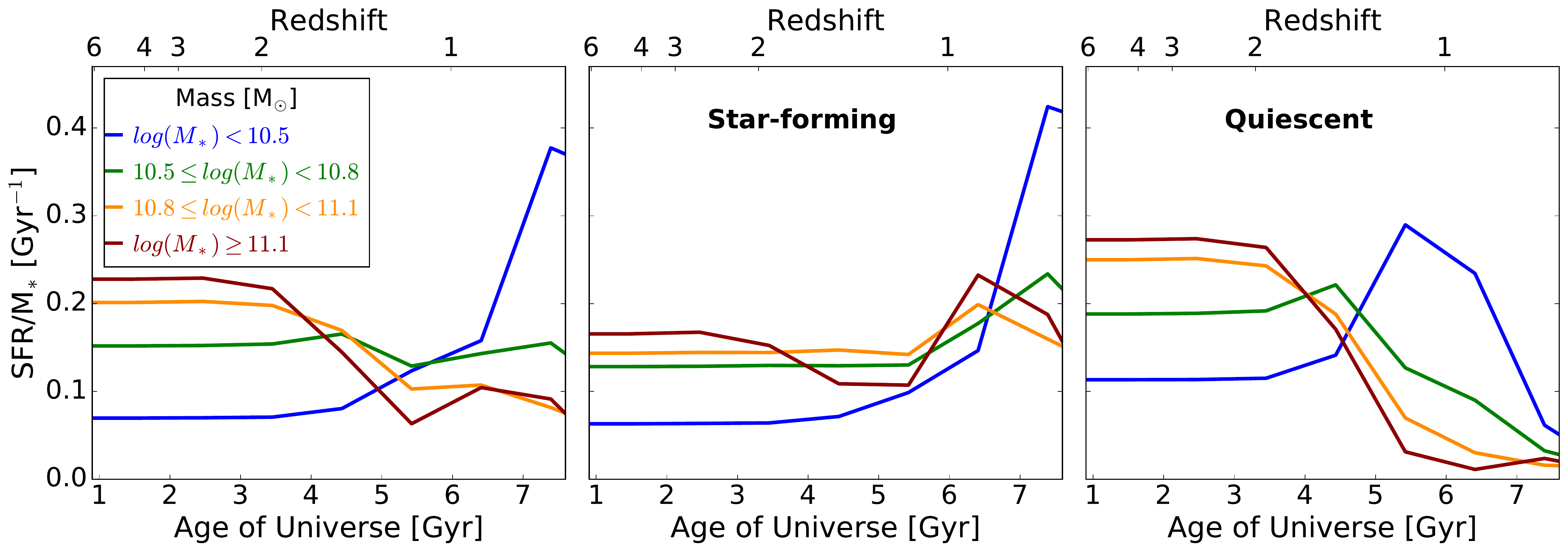}{0.95\textwidth}{(b)}}
\label{fig:sfr_lb_mbins_b}
\caption{Ensemble-averaged SFHs of LEGA-C galaxies, normalised by stellar mass and separated into various \deleted{velocity dispersion} \added{\sigs\,} (top) and \deleted{stellar mass} \added{\mfit,} bins (bottom). The histories are divided into the star-forming and quiescent populations in the middle and right panels, respectively. The stellar content in massive galaxies formed earlier and faster, regardless of current SF activity.}
\label{fig:sfr_lb_mbins}
\end{figure*}

\begin{figure*}[t]
\centering
\includegraphics[width=0.8\textwidth]{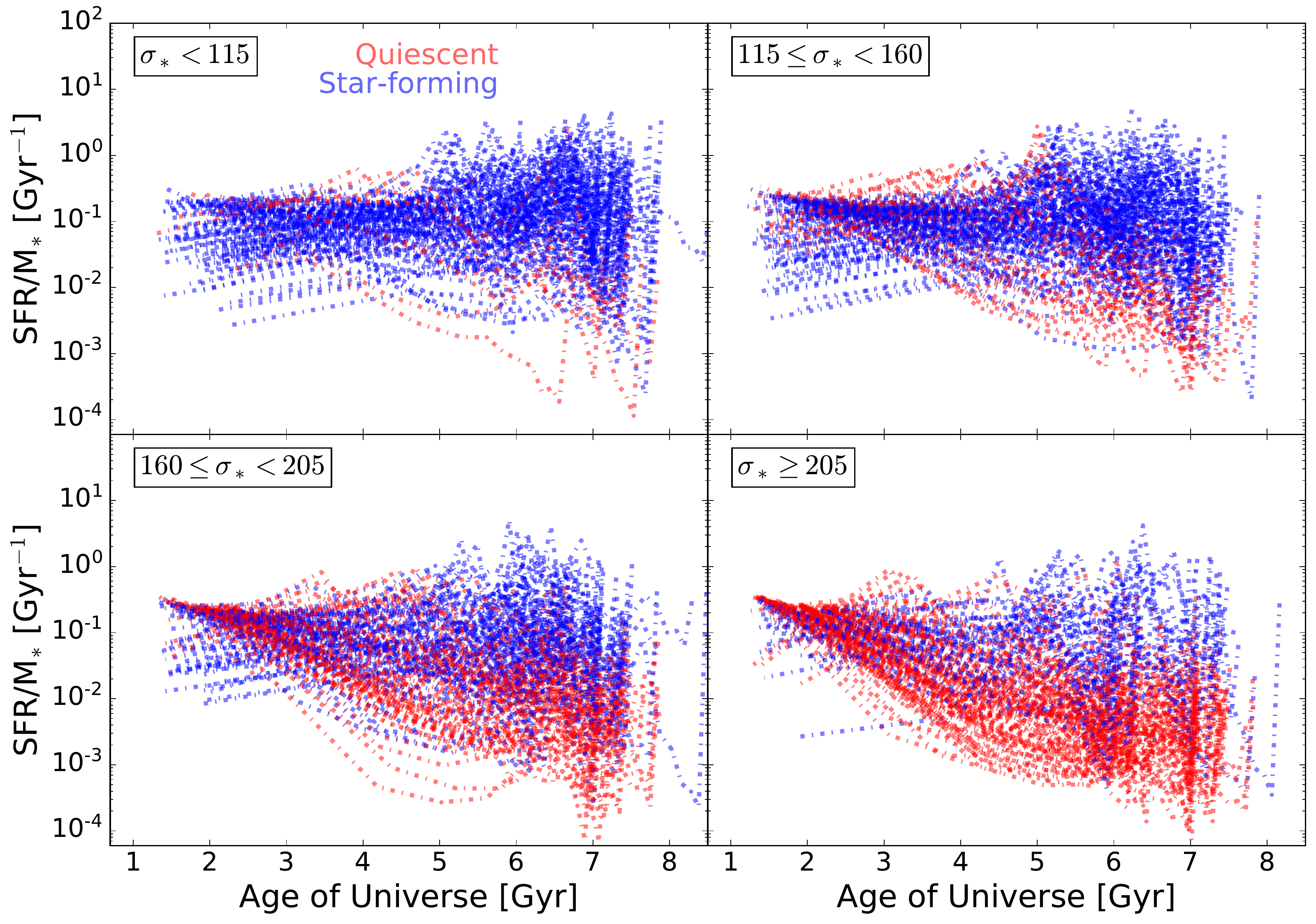}
\caption{SFHs of the LEGA-C sample (normalised by stellar mass) as a function of the age of the Universe separated into four \deleted{velocity dispersion} \added\sigs{\,} bins indicated by the labels. The colours differentiate between the star-forming and quiescent populations at the observed redshift.}
\label{fig:sfr_q_sf_vdisp}
\end{figure*}

\section{SFHs of the Galaxy Population}
\label{sec:mr}

\added{\subsection{Correlations between Age, \mfit and \sigs}}
\label{sec:corr}

\added{Figure \ref{fig:age_vs_v_m} shows \amw\, } as a function of \deleted{M$_{*,FIT}$ colour-coded by \amw\,(} \added{stellar mass (\mfit\,, } left panel) and \deleted{divided } \added{stellar velocity dispersion (\sigs\,, right panel) colour-coded } by current SF activity, i.e. whether the galaxy is quiescent (log(sSFR$_{UV+IR}$[Gyr$^{-1}]) <-1$) or star-forming. Selecting quiescent galaxies by their U-V and V-J colors would result in similar trends. \added{\amw\, generally correlates more strongly with \mfit\, than \sigs. However, there is a \sigs\, threshold above which galaxies are almost exclusively old and quiescent: galaxies with \sigs\,$>200-250$\,km\,s$^{-1}$ and \amw\,$<4$\,Gyrs are very rare. Such a clear threshold does not exist for \mfit: high-mass galaxies (\mfit\,$\gtrsim10^{11}$\,M$_{\odot}$) show a variety of ages.}

\added{To further illustrate these trends, we show \sigs\, as a function of \mfit\, colour-coded by \amw\, (left panel) and divided by current SF activity (middle and right panels) in Figure \ref{fig:vdisp_vs_mass}.} There is a discernible separation between old \added{($>4$\,Gyrs)} and young \added{($<4$\,Gyrs) galaxies at a velocity dispersion of \sigs\,$\sim170$\,km\,s$^{-1}$.}\deleted{with velocity dispersion ($\sigma_*\sim170$\,km\,s$^{-1}$) that spans a broad range of M$_{*,FIT}$.}  \added{Taken together with the trends seen in Figure \ref{fig:age_vs_v_m}, we can conclude that \sigs\,$>250$\,km\,s$^{-1}$  is a sufficient requirement for having an old age and \sigs\,$\sim170$\,km\,s$^{-1}$ is a necessary requirement for old age.} This extends the properties of present-day early-type galaxies, for which a correlation between \deleted{structure} \added{velocity dispersion (and closely related quantities such as surface mass density and central mass density)} and stellar age has been shown to be more fundamental than age trends with stellar mass \citep{kauffmann2003,vanderwel2009,graves2009}, to higher redshift. Our results also extend the widely reported correlation between velocity dispersion \deleted{(and closely related quantities such as} \added{(as well as} surface mass density and central mass density) and \deleted{SFR} \added{SF activity} \citep[e.g.,][]{franx2008,mosleh2017,barro2017} to an underlying correlation with overall stellar age.

The \added{scaling} relation between \sigs\, and black hole (BH) mass implies that large BH mass is correlated with early SF and the ceasing thereof. Such a scenario is supported by the direct correlation between BH mass and SF activity \citep[e.g.,][]{Terrazas2016} and the large fraction of radio AGN among galaxies with large velocity dispersions both at low and high redshift \citep[e.g.,][]{best2005, barisic2017}. 

It is interesting to note that \deleted{\amw\, does not correlate with \sigs\,} \added{the correlation between \amw\, and \sigs\, seen in Figure \ref{fig:vdisp_vs_mass} is significantly weakened} after dividing the population by current SF activity\deleted{ (Figure \ref{fig:vdisp_vs_mass})}. Instead, \added{for} the star-forming population\added{, galaxy age is better} \deleted{is more} correlated with \mfit\,\deleted{ and the quiescent population has no preference for either parameter} \added{(also seen in Figure \ref{fig:age_vs_v_m})}. \deleted{The stronger correlation between \amw\, and \mfit\, for star-forming galaxies indicates that when SF starts in a galaxy, its SFH is to first order constant such that \mfit\, simply traces age (SFH). On the other hand, most quiescent galaxies in our sample are very old (see Figure \ref{fig:hist} and \ref{fig:fit1}), therefore, they form most of their stars in our oldest age bin, which potentially hides an existing correlation between \sigs\, or \mfit\, with SFH. We still don't fully resolve the SFHs of these galaxies; whether that is the limitation of the data or a choice of age bins remains to be seen.} \added{A straightforward interpretation is that when galaxies are growing rapidly through SF--that is, when they are located on or near the SF `Main Sequence'--then M$_*$ mostly traces how long this main SF phase has lasted so far. In other words, M$_*$ simply traces the build-up of the stellar population over time, while \sigs\, is related to the end of this main SF phase, i.e. to the regulation and cessation of SF, presumably through AGN feedback.}

\subsection{Evolution of the average SFHs}
\label{sec:sfh1}

The \deleted{(completeness corrected) } average SFHs of galaxies, normalised by stellar mass, as a function of $\sigma_*$ \deleted{(upper panel)} and \mfit\, \deleted{(lower panel)} are shown in Figure \ref{fig:sfr_lb_mbins}(a) and (b), respectively. The \added{average SFHs were corrected for completeness by weighing each galaxy by a completeness correction factor to create a volume-limited quantity 
\cite[see ][]{wu2018}. The } population is divided by its current star-formation activity in order to disentangle the effects from these two populations as well as compare them. The velocity dispersion and mass ranges and were selected such that there were enough galaxies in each bin ($\geq10$), in both quiescent and star-forming galaxies. These relations are used to determine whether $z\sim1$ galaxies also show a downsizing trend in their SFHs, as many studies have pointed to using local galaxies. However, the SFHs seen at $z\sim1$ would not be resolved at $z\sim0.1$, as the stellar populations would be too old.

On average, high-$\sigma_*$ galaxies ($\sigma_* \geq 170$\,km\,s$^{-1}$) had higher SFRs at earlier epochs which started to decline rapidly, at a rate that increases with $\sigma_*$ and stellar mass, when the Universe was $\sim3$\,Gyrs old. Most galaxies with lower velocity dispersions ($\sigma_* < 170$\,km\,s$^{-1}$) gradually build their stellar mass as the Universe evolves; however, the SFR of a minority, i.e. the quiescent population, began to decline when the Universe was $\sim5$\,Gyrs old. Higher-mass star-forming galaxies (\mfit\,$\geq10^{10.5} M_{\odot}$) have SFHs that are consistent with constant star-formation with time, while the lower mass galaxies (\mfit\,$<10^{10.5}$) still have rising SFRs. The star-forming population is still undergoing its main formation phase. The SFH trend is clear with \mfit\, and not \sigs\, for the star-forming population, which extends from \mfit\, being better correlated with SFHs for star-forming galaxies as discussed in Section \ref{sec:corr} (see Figure \added{\ref{fig:age_vs_v_m} and} \ref{fig:vdisp_vs_mass}).

Figure \ref{fig:sfr_lb_mbins} reveals that, on average, most galaxies in the sample were forming stars quite early on; however, the SFRs were systematically higher and the eventual decline systematically more rapid with increasing \sigs\, (\mfit\, for the star-forming population). This is clear evidence for the top-down scenario; where galaxies downsize in their star formation with time (more massive galaxies have older stars). This is seen in the overall population, and more strongly so in the quiescent population. While this result is in alignment with previous studies for the local universe \citep[e.g. ][]{juneau2005, thomas2005, tojeiro2009, mcdermid2015, ibarra2016}, our work establishes this trend at $z\sim1$ (half the current age of the Universe) using \deleted{spectroscopy}\added{using full spectrum fitting. \cite{wu2018}'s study of the D$_n$4000 and H$\delta$ spectral features also support the downsizing scenario.}

\subsection{The variety of SFHs}
\label{sec:sfh2}

In Figure \ref{fig:sfr_q_sf_vdisp}, we show all the stellar mass normalised SFHs in the LEGA-C sample, separated into four velocity dispersion bins and divided into the quiescent and star-forming population (at the observed redshift) as defined in Section \ref{sec:corr}. This reveals the large scatter in the SFHs at fixed mass, in addition to discerning the differences in the histories based on the current star-formation activity of the galaxies.

The SFHs of quiescent galaxies peaked early on in the Universe and thereafter, their activity generally decreases with time; while star-forming galaxies gradually grow in SFR, which peaked at later epochs. The quiescent population has consistently higher SFRs at early epochs, whereas its star-forming counterpart has higher SFRs at later epochs. This indicates that star-forming galaxies aggregate their mass slower than the quiescent population. The dominance of the quiescent population increases from low to high-mass galaxies, and vice versa for the star-forming population.

The SFRs of low mass galaxies ($\sigma_* < 115$\,km\,s$^{-1}$) have been gradually increasing, with large scatter at all epochs. These galaxies are currently undergoing the main stages of their star formation. Note that the lowest dispersion bin suffers from incompleteness, due to the survey sample selection approach. K-band quiescent galaxies are fainter than equally massive star-forming galaxies which causes an under-representation in the LEGA-C sample. However, it is well known that low-mass star-forming galaxies outnumber quiescent galaxies of the same mass; therefore, the SFHs in Figure \ref{fig:sfr_q_sf_vdisp} can be considered as illustrative.

The quiescent and star-forming populations are more evenly distributed \deleted{in } \added{(in number and variation of SFHs) in } the intermediate-$\sigma_*$ regime (between $160$ and $205$\,km\,s$^{-1}$) \deleted{, and their SFHs have less scatter overall. On the other hand, the } \added{while the } high-$\sigma_*$ population ($\sigma_* \geq 205$\,km\,s$^{-1}$) is dominated by quiescent galaxies. The \deleted{scatter in the dispersion bins $\geq160$\,km\,s$^{-1}$ was low when the Universe was young ($>10$\,Gyrs ago) regardless of whether the galaxy is currently forming stars or not; it increased towards lower redshifts, with the highest dispersion bin ($\sigma_* \geq 205$\,km\,s$^{-1}$) having larger scatter at lower lookback times ($<9$\,Gyrs ago). Taking into account that the difference in scatter could be due to the size of the age bins, by averaging out the star formation at lookback\,$<10$\,Gyrs to mimic a larger age bin, we still measured larger scatter at lower lookback times. The star-formation activity of star-forming galaxies in this mass regime peaked \,$\sim8.5$\,Gyrs ago. The } disparity between the SFHs of the quiescent and star-forming populations \added{in the high-$\sigma_*$ regime } indicates that galaxies `remember' their past. There is a strong coherence among the SFHs of quiescent and star-forming galaxies, respectively. This behaviour extends to the peak of cosmic SF activity at $z\sim$\,2-3. This implies that SF activity at the moment of observation is strongly correlated with the SF activity $\sim3$\,Gyrs prior. The results of this work indicate that many evolutionary paths can lead to galaxies at a given velocity dispersion. This illustrates the difficulty of connecting progenitor and descendant populations at different cosmic epochs.

\subsection{Comparisons to Literature Measurements}
\label{sec:sfh3}

As stated in Section \ref{sec:sfh1}, the deconstructed SFHs in this study support the galaxy downsizing scenario which has long been studied \deleted{. In terms of stellar populations, many of these studies involved the use of fossil record methods on SDSS spectra of local galaxies. Downsizing has also been seen in other studies, such as studies by \cite{cimatti2006}, who corrected luminosity function data of early-type galaxies by adopting the empirical luminosity dimming rate derived from the evolution of the Fundamental Plane of field and cluster massive early-type galaxies, as well as \cite{leitner2012}, who derived the average growth of stellar mass in local star-forming galaxies using a Main Sequence Integration approach. \cite{leitner2012} found} \added{(see Section \ref{sec:intro}). \cite{leitner2012}'s finding } that star-forming galaxies formed only $\sim15$\,\% of their mass before $z=$\,1-2 (mass dependent)\deleted{which suggests }\added{, suggesting } that present-day star-forming galaxies are not the descendants of massive star-forming galaxies at $z>2$\deleted{. This notion }\added{, } is in line with our results since the peak in star formation occurs after $z<1.5$ for almost all star-forming galaxies in the LEGA-C sample.

Intermediate-redshift stellar population studies are sparse, due to the high S/N required to undertake such studies \added{(see Section \ref{sec:intro})}. Pertaining to this work, there are a few studies we can draw comparisons from, viz. \deleted{
\cite{schiavon2006}, } \cite{choi2014} and \cite{gallazzi2014}. \deleted{
\cite{choi2014} 
used a spectral fitting algorithm on SDSS quiescent galaxies in the redshift range $0.1<z<0.7$ to investigate the evolution of stellar ages over time as a function of mass. They found that the increase in stellar ages with time for massive galaxies is consistent with passive evolution since $z=0.7$. The same conclusion was reached  for quiescent galaxies by 
\cite{gallazzi2014} 
who characterised the stellar age-stellar mass and stellar metallicity-stellar mass relations for a sample of $\sim70$ galaxies at $z\sim0.7$. These results } \added{Measurements by 
\cite{choi2014} 
and 
\cite{gallazzi2014} 
indicating that passive galaxies have ages consistent with mostly passive evolution } are also in alignment with this study as the reconstructed SFHs indicate that galaxies stay quiescent, barring some histories that showed low-level star formation after quiescence. \cite{gallazzi2014} \deleted{also measured a flattening of the stellar metallicity-stellar mass relation towards solar metallicity (M$_*>3\times10^{10}$\,M$_{\odot}$), with significant scatter. This incited our choice to use solar metallicity in the fitting algorithm, as we found no significant differences in the goodness of the fits, in terms of the normalised $\chi^{2}$, when using either sub-solar or super-solar metallicity instead. 
\cite{gallazzi2014} 
} reported an average lighted-weighted age of $\sim5$\,Gyrs for a $4\times10^{10}$\,M$_{\odot}$ galaxy, consistent with our value of $4.8$\,Gyrs, for a galaxy of the same mass.

\cite{diemer2017} tested \cite{gladders2013}'s hypothesis that the SFHs of individual galaxies are characterised by a log-normal function in time, which implies a slow decline in SFRs rather than rapid quenching. They did this by comparing the log-normal parameter space of total stellar mass, peak time, and full width at half maximum of simulated galaxies from Illustris \citep{vogelsberger2014} and \cite{gladders2013}, as well as \cite{pacifici2016}'s derived SFHs of a sample of quiescent galaxies using a large library of computed theoretical SFHs. They found good agreement for all three studies, however, Illustris predicted more extended SFHs on average. LEGA-C galaxies support the slow-quenching picture of galaxy evolution as \cite{gladders2013} have suggested, with a rate of decline that is mass dependent as we have seen. More comparisons will be performed in later papers.

\section{Summary}
\label{sec:summ}

We have reconstructed the SFHs of galaxies in the current LEGA-C sample, which contains 678 primary sample galaxies with S/N\,$\sim20$\,\AA$^{-1}$ in the redshift range  $0.6<z<1$. We have done this by implementing an algorithm to fit flexible SFHs to the full spectrum, using \textit{FSPS} and \textit{emcee}. The galaxy spectra were fit to a linear combination of a defined set of 12 CSPs, with solar metallicity and constant star formation within the time interval of the templates. In 90\% of the cases the algorithm produced good fits based on the normalised $\chi^2$ values. We found a wide variety of SFHs, although 60\% of the galaxies have \amw\,$>3$\,Gyrs by the time we observe them (Figures \ref{fig:hist} and \ref{fig:fit1}). However, we note that age estimates from spectral fits experience increasing degeneracy of spectral features as the stellar populations age. Most of the old galaxies (\amw\,$\gtrsim3$\,Gyrs) had very low SFRs early on ($\gtrsim6$\,Gyrs after the Big Bang, Figure \ref{fig:fit1}). However, some exhibit subsequent peaks in star formation, which could be an indication of rejuvenated star formation, or a merger with a younger population. However, the mass formed from this more recent star formation activity is only about 10\% of the mass formed throughout the galaxies' histories. The median \alw, \amw\, and \mfit\, were found to be $1.2$\,Gyrs, $3.8$\,Gyrs and $10^{10.8}$\,M$_{\odot}$, respectively.

The main objective of this work was to investigate how \deleted{the } \added{our } reconstructed SFHs behave as a function of \added{stellar mass, stellar } velocity dispersion and \deleted{stellar mass} \added{star-formation (SF) activity}, as well as \deleted{scatter } \added{the variation } they show at fixed velocity dispersion. We found that galaxies at $z\sim1$ have similar trends in their SFHs compared to local galaxies, i.e. the stellar content in massive galaxies formed earlier and faster (Figure \ref{fig:sfr_lb_mbins}). This top-down scenario is a known trend from fossil record inferences using SDSS spectra; however, in this study, it is shown for $z\sim1$ galaxies for the first time  \added{using full spectrum fitting}. We found that the scatter \deleted{in SFHs of high-dispersion galaxies ($\sigma_* \geq 205$\,km\,s$^{-1}$) was low when the Universe was young ($>10$\,Gyrs ago), regardless of the current star-formation activity (Figure \ref{fig:sfr_q_sf_vdisp}). However, the scatter } between the quiescent and star-forming populations increases towards lower redshift (\deleted{lookback\,$<9$\,Gyrs}\added{Figure \ref{fig:sfr_q_sf_vdisp}}), which indicates that current SF activity is strongly correlated with past SF activity. High-dispersion \deleted{star-forming } \added{quiescent } galaxies had their star formation peak \deleted{around $8.5$\,Gyrs ago, while quiescent galaxies peaked earlier, } \added{early, } $>9.5$\,Gyrs ago, and exhibit decreasing SFRs throughout the rest of their history, for the most part. We found that the lowest dispersion galaxies in our sample are undergoing the main stage of their star formation as we observe them ($7$\,Gyrs ago).

The results of the spectral fits were used to measure a number of galaxy properties, viz. ages (\alw, \amw, \deleted{\amed } \added{\amw, } etc.) and stellar mass, in order to test the model by investigating how these properties relate to one another as well as other properties measured from the galaxy spectra, e.g. velocity dispersion, $H{\delta}$, D$_n$4000, etc. We showed that galaxies evolve from the top-left to the bottom-right of the EW(H$\delta$)-D$_n$4000 plane as they age, as would be expected (Figure \ref{fig:hdd4000}).\deleted{ The stellar masses measured from our model agreed well with those measured from FAST, which further supported the results of our spectral fits (Figure \ref{fig:masscomp}).}

Recovering the full SFHs of intermediate-redshift galaxies opens up a multitude of avenues of research. In this work we have shown the clear differences between the SFHs of quiescent and star-forming galaxies and how these SFHs are scattered at fixed velocity dispersion. We have also shown that velocity dispersion is a better indicator of the age and current SF activity of galaxies as a whole than stellar mass, while stellar mass is a better indicator of the age of star-forming galaxies (Figure \added{\ref{fig:age_vs_v_m} and } \ref{fig:vdisp_vs_mass}). \deleted{With SFHs at these redshifts, we can } \added{In future studies, we will use the reconstructed SFHs to } constrain the quenching speed and rate, as well as investigate the relationship between galactic structure and SFHs\deleted{(this will be done in future papers)}. These constraints will become valuable for future surveys like JWST that will be investigating the properties of galaxies beyond $z\sim2$, and will need $z\sim1$ measurements as a benchmark to connect those populations to the local Universe.

\section{Acknowledgements }
\label{sec:ack}

Based on observations made with ESO Telescopes at the La Silla Paranal Observatory under programme ID 194-A.2005 (The LEGA-C Public Spectroscopy Survey).  PC gratefully acknowledge financial support through a DAAD-Stipendium. This project has received funding from the European Research Council (ERC) under the European Union's Horizon 2020 research and innovation programme (grant agreement No. 683184). KN and CS acknowledge support from the Deutsche Forschungsemeinschaft (GZ: WE 4755/4-1). We gratefully acknowledge the NWO Spinoza grant.







\bibliography{sfh_legac_apj_changes}


\listofchanges

\end{document}